\documentstyle[aps,prd,epsfig]{revtex}

\begin{document}
\tightenlines
\title{Trilinear Neutral Gauge Boson Couplings in Effective Theories}
\author{F. Larios}
\address{Departamento de F\'\i sica Aplicada, CINVESTAV-M\' erida,
Apartado Postal 73, 91310, M\' erida, Yucat\' an, M\' exico}
\author{M. A. P\'erez and G. Tavares-Velasco}
\address{Departamento de F\'\i sica, CINVESTAV, Apartado Postal 14-740,
07000, M\' exico, D. F., M\' exico}
\author{J. J. Toscano}
\address{Facultad de Ciencias F\'\i sico Matem\' aticas, Benem\' erita
Universidad Aut\' onoma de Puebla, Apartado Postal 1152, 72000, Puebla,
Pue., M\' exico}

\date{\today}
\maketitle

\begin{abstract}

We list all the lowest dimension effective operators inducing off-shell
trilinear neutral gauge boson couplings $Z Z \gamma$, $Z \gamma \gamma$,
and $Z Z Z$ within the effective Lagrangian approach, both in the linear
and nonlinear realizations of the $\mathrm{SU(2)_{L} \times U(1)_{Y}}$
gauge symmetry. In the linear scenario we find that these couplings can be
generated only by dimension eight operators necessarily including the
Higgs boson field, whereas in the nonlinear case they are induced by
dimension six operators. We consider the impact of these couplings on some
precision measurements such as the magnetic and electric dipole moments of
fermions, as well as the $Z$ boson rare decay $Z \to \nu \overline{\nu}
\gamma$. If the underlying new physics is of a decoupling nature, it is
not expected that trilinear neutral gauge boson couplings may affect
considerably any of these observables.  On the contrary, it is just in the
nonlinear scenario where these couplings have the more promising prospects
of being perceptible through high precision experiments. 

\end{abstract}
\draft

\section{Introduction}
\label{intro}

The present agreement between experimental data and the standard model
(SM) suggests that the energy scale $\Lambda$ associated with any new
physics should be large compared with the electroweak scale $v=(\sqrt 2
G_F)^{1/2}= 246$ GeV. To infer the existence of new particles as heavy as
$\Lambda$ through their virtual effects, effective Lagrangian (EL) 
techniques have been extensively used to study quantities which are
forbidden or highly suppressed within the SM \cite{bola,sza,maya}. Among
these quantities, self-couplings of electroweak gauge bosons constitute a
sensitive probe of nonstandard interactions \cite{ellison}. Experimental
bounds on possible anomalous $W^+W^-Z(\gamma)$ couplings have reached an
accuracy of the few percent level in both hadronic and leptonic colliders
\cite{wwz,lep2}, but the situation looks less promising for anomalous
$ZZZ, ZZ\gamma$, and $Z\gamma \gamma$ couplings
\cite{zzgamma}\footnote{Throughout this work we consider the general case
of off-shell bosons, unless stated otherwise, but they will be denoted by
$V$ rather than $V^*$.}. Unlike $W^+W^-Z(\gamma)$ couplings, trilinear
neutral gauge boson couplings (TNGBC) vanish when the three bosons are
real. Another interesting peculiarity of TNGBC is that they must be
induced by loop effects in any renormalizable theory since they cannot
possess a renormalizable structure. In the SM, TNGBC are generated at
one-loop level by fermion triangles \cite{rb}, being very suppressed even
in the presence of a fourth fermion family \cite{hernandez}. There follows
that it is convenient to carry out a model independent study of TNGBC
using the EL method to parametrize any anomalous contribution. Within this
approach, there are two well motivated schemes to parametrize virtual
effects of physics beyond the Fermi scale via effective operators
involving only SM fields, namely the linear and the nonlinear
realizations.

In the linear realization or decoupling scenario it is assumed that the
light spectrum of particles, which fill out multiplets of the electroweak
$ \mathrm{SU(2)_{L} \times U(1)_{Y}} $ gauge group, includes at least the
physical Higgs boson of the SM. Because of the decoupling theorem, virtual
effects of heavy physics cannot affect low energy processes dramatically. 
Nonetheless, any new effect, in spite of its smallness, may have
significant effects on the couplings which are absent or highly suppressed
within the SM. Starting from the SM fields and assuming lepton and baryon
number conservation, there is no way to construct any odd dimension
operator respecting the linearly realized $ \mathrm{SU(2)_{L} \times
U(1)_{Y}} $ symmetry. As for dimension six, operators of this class were
comprehensively studied in \cite{bw}. It was shown that there are 84
independent dimension six operators.

In the case of the nonlinear realization or nondecoupling scenario, the
parametrization of new physics effects arises when it is assumed that the
Higgs bosons are very heavy or do not exist at all. The scalar sector is
comprised only by Goldstone bosons, which transform nonlinearly under the
$ \mathrm{SU(2)_{L} \times U(1)_{Y}} $ group. It is also possible to
introduce light scalar fields in this parametrization, but they cannot be
recognized as Higgs bosons since such fields do not couple to the
remaining light particles as dictated by the Higgs mechanics \cite{b1}. 
Since the low energy theory is nonrenormalizable under Dyson prescription,
heavy physics does not decouple from the low energy processes. We may
think of this scenario as the one in which the EL parametrizes unknown
physics which would not obey the Higgs mechanism. In this case, the most
important operators are the ones which induce the masses of the $W$ and
$Z$ gauge bosons, prescribing also the general structure of the
$W^+W^-Z(\gamma)$ couplings \cite{nlr}. These operators have dimension two
and four.

At the lowest order, anomalous $W^+W^-Z(\gamma)$ couplings are induced by
dimension six operators in the decoupling scenario. In the nonlinear
scheme, they receive contributions from dimension four operators. In
contrast, at the lowest order, TNGBC are induced by dimension eight
operators in the linear realization and by dimension six operators in the
nonlinear one. In the latter case there are also some dimension four
operators which give rise to the $ZZZ$ coupling, but they are proportional
to the scalar part of the $Z$ boson ($\partial_\mu Z^\mu$). It can be
shown that such operators may be eliminated by means of a transformation
which leaves invariant the $S$-matrix \cite{ew}. Consequently, any
anomalous contribution to TNGBC is expected to be more supppresed than
those inducing nonstandard $W^+W^-Z(\gamma)$ couplings.  It must be
stressed, however, that any potential effect must be carefully examined as
it may constitute a clear evidence of new physics. 

The structure of TNGBC has already been studied in the context of
effective theories, initially at the level of vertex functions
\cite{hpzh}. However, in this approach it was considered the case where
two particles are real and just one is virtual. It is only very recently
that the analysis of the off-sell vertices has been done under the
$\mathrm{U(1)_{em}}$ gauge invariant framework, including the study of the
respective EL. By invoking Bose symmetry, Lorentz covariance, and
electromagnetic gauge invariance, the most general structures inducing
TNGBC with three off-shell neutral bosons were constructed \cite{r}. As
was shown in \cite{b2}, the $\mathrm{U(1)_{em}}$ gauge invariant framework
is equivalent to the nonlinearly realized $ \mathrm{SU(2)_{L} \times
U(1)_{Y}} $ invariant case. Such an equivalence is explicit in the unitary
gauge. The choice of using either framework is only a matter of
convenience.  In particular, the nonlinear scheme is convenient in working
out loop calculations, as the presence of Goldstone bosons allows to
quantize the theory with the aid of a renormalizable $R_\xi$ gauge. 

It is clear that a comprehensive study of TNGBC must include both the
linear and the nonlinear schemes. To our knowledge the former has never
been studied before. One of the aims of the present paper is to present a
complete list of the effective operators which induce TNGBC at the lowest
order in both realizations of the $ \mathrm{SU(2)_{L} \times U(1)_{Y}} $
gauge symmetry. Not all the operators that can be constructed respecting
the Lorentz and electroweak symmetries are independent since a certain
class of general transformations allows to rule out some of them without
affecting the $S$-matrix elements \cite{ccwz}. In the course of our
classification we have found operators with terms containing higher
derivatives which resemble the covariant structure of the equations of
motion; there are also operators with terms which are proportional to the
scalar part of the $Z$ boson $(\partial_\mu Z^\mu)$. It has been shown in
\cite{ew,a} that both types of structures can be eliminated in favor of
other operators already present in the effective Lagrangian. Such a
procedure is only valid at first order in the unknown effective parameters
of the theory as any effective Lagrangian is assumed to describe the
effects of well-behaved new physics just in this approximation. 
Consequently, after performing the required transformation, the equations
of motions can be used to eliminate any redundant structure, expressing
the respective operator in terms of other ones. This whole procedure does
not affect the $S$-matrix elements. In order to present all the
independent operators, we will classify them according to the following
criterion: those which can not be reduced using the equations of motion
will be referred to as irreducible, the remaining ones will be referred to
as reducible. 

After classifying the operators, our paper will be concerned with the
sensitivity of some precision experiments to new physics effects arising
from TNGBC.  Although persuasive theoretical arguments indicate that
trilinear gauge boson couplings are not expected to be larger than the one
percent \cite{rghm,aew}, the Large Hadron Collider (LHC) as well as the
planned Next Linear Collider (NLC) are expected to constrain them at a
level of $10^{-4}$--$10^{-6}$ \cite{ellison,glr}. As long as TNGBC are
concerned, the size of their effects will be suppressed by powers of
$(v/\Lambda)^4$ and $(v/\Lambda)^2$ in the linear and the nonlinear
scenarios, respectively. We will examine whether some high precision
measurements may lead to any reasonable bound on these couplings. The
anomalous $W^+W^-\gamma (Z)$ couplings have been constrained from a global
analysis of the LEP/SLC observables at the $Z$ pole \cite{sza}. To draw
any inference about the size of TNGBC we will consider the muon $g-2$
value, the known limit on the electric dipole moment (EDM) of the
electron, and the current limit on the rare decay $Z \to \nu
\overline{\nu} \gamma$. 

Our paper is organized as follows. All the lowest dimension operators that
generate TNGBC in the linear scheme are presented in Sec. II, following
the classification criterion already explained. Besides, the respective
Lagrangians and vertex functions are shown explicitly. In Sec. III, a
similar analysis within the nonlinear scenario is presented. Sec. IV is
devoted to examine the possibility of obtaining constraints on the
couplings out of high precision experiments. Finally, the paper is closed
with some concluding remarks in Sec. V. 

\section{The decoupling scenario} \label{lc}

This section focuses on the itemization of all the lowest dimension
operators that generate at least one of the couplings $ZZZ$, $ZZ \gamma$
or $Z \gamma \gamma$ within the linear realization of the $
\mathrm{SU(2)_{L} \times U(1)_{Y}} $ electroweak group. To construct a
basis of independent operators with a given dimension, we must consider
some aspects concerning the independence of the $S$-matrix under a wide
class of transformations which leave it invariant \cite{ccwz}. For
instance, it was shown in \cite{a} that some operators, which consist of a
piece containing higher derivatives, can be eliminated in favor of others
by using a specific transformation, leaving unchanged the $S$-matrix
elements at any order of perturbation theory. Another situation arises
when an operator is proportional to the scalar part of the $Z$ boson. 
While the latter kind of structures give vanishing contributions when the
$Z$ boson is on mass shell or is virtual but couples to light fermions,
the situation is not the same in the case of the top quark. In this
respect, this kind of operators can also be eliminated by performing a
transformation which does not alter the $S$-matrix elements \cite{ew}.  It
must be noted that both transformations are equivalent to applying the
equations of motion. Beside these considerations, we have made a
systematic use of integration by parts to rule out any operator related to
others through a surface term.  Consequently, we will catalog the
operators inducing TNGBC as reducible or irreducible. 

Any $ \mathrm{SU(2)_{L} \times U(1)_{Y}} $ invariant involving only
bosonic fields can be constructed out of the covariant structures $B_{\mu
\nu}$, ${\bf W}_{\mu \nu}=\frac{1}{2}\sigma^i W^i_{\mu \nu}$, $\Phi$, and
$D_\mu \Phi$, where the covariant derivative is defined as
$D_\mu=\partial_\mu- ig{\bf W}_\mu-ig'B_\mu$, and $\Phi$ is the Higgs
doublet. Using these basic structures, we can built the following $
\mathrm{SU(2)_{L} \times U(1)_{Y}} $ invariant and Lorentz covariant
structures of dimension two through five

\begin{equation}
\label{builbloc}
B_{\mu \nu}, \, \Phi^{\dag}\Phi, \, \Phi^{\dag}D_\mu\Phi, \,
\Phi^{\dag}{\bf W}_{\mu \nu}\Phi, \, B_{\mu \nu}B^{\lambda \rho}, \,
{\rm{Tr}}[{\bf W}_{\mu \nu}{\bf W}^{\lambda \rho}], \, \Phi^{\dag}(D_\mu
D_\nu+D_\nu D_\mu)\Phi, \, \Phi^{\dag}{\bf W}_{\mu \nu}D_\lambda\Phi. 
\end{equation}

\noindent Note that another set of $ \mathrm{SU(2)_{L} \times U(1)_{Y}} $
invariant and Lorentz covariant structures can be generated by operating
with the ordinary derivative on these expressions. Any nonrenormalizable
bosonic operator can be built by choosing the appropriate combinations of
these structures to form Lorentz scalars. The ordinary derivative can act
on the last expressions in several ways, but the contractions
$\partial^\mu B_{\mu \nu}$ and $\partial^\mu (\Phi^\dag D_\mu \Phi)$,
being proportional to the scalar part of the $Z$ boson, are special
because in both cases we can use the equations of motion to eliminate the
resulting operator. 

Let us now discuss the general Lorentz structure of TNGBC. The lowest
dimension operators which can be assembled out of the basic structures
have dimension six \cite{bw}. It is easy to see that no dimension six
operator induce TNGBC, which unavoidably leads to search for eight
dimension operators. In principle, the combination which can give rise to
TNGBC may involve the 4-vectors $A_\mu$ and $Z_\mu$, together with the
antisymmetric tensors $F_{\mu \nu}=\partial_\mu A_\nu-\partial_\nu A_\mu$
and $Z_{\mu \nu}=\partial_\mu Z_\nu-\partial_\nu Z_\mu$.  Owing to
$\mathrm{U(1)_{em}}$ gauge symmetry, the electromagnetic field can only
appear as $A_\mu$ through the respective covariant derivative, which
operates on charged fields only. Therefore, the photon must appear in any
term through the tensor field $F_{\mu \nu}$. Due to the antisymmetry of
the $F_{\mu \nu}$ and $Z_{\mu \nu}$ tensors, it is not possible to
generate TNGBC using only these structures: it would be necessary to have
at our disposal three antisymmetric tensors. There follows the absence of
the $\gamma \gamma \gamma$ vertex in this gauge invariant scheme.

To construct the $ZZZ$, $ZZ\gamma$, and $Z\gamma \gamma$ vertices, we must
use at least a $Z$ boson in the $Z_\mu$ form, which is allowed because
this field couples to neutral fields. The 4-vector $Z_\mu$ is contained in
the covariant derivative, which in the bosonic sector operates only on the
Higgs doublet. As a consequence, the Higgs mechanism plays a special role
in this type of couplings. In particular, the Higgs presence increases the
dimension at which the operators can be generated in comparison to the
nonlinear case, where this field is absent. The $Z$ boson may appear
through the combinations $Z_{\lambda \rho} Z_{\mu \nu}$, $Z_\lambda Z_{\mu
\nu}$, $Z_\mu Z_\nu$, and $Z_\mu$. The building blocks necessary to
construct these couplings are $\Phi^{\dag}D_\mu\Phi$, $\Phi^{\dag}(D_\mu
D_\nu+D_\nu D_\mu)\Phi$, and $\Phi^{\dag}{\bf W}_{\mu \nu}D_\lambda\Phi$,
which, after spontaneous symmetry breaking (SSB), induce the structures
$Z_\mu$, $Z_\mu Z_\nu$, and $Z_{\mu \nu} (F_{\mu \nu})Z_\lambda$,
respectively. The irreducible operators may contribute to a given physical
process through the specific structure of TNGBC, while the reducible ones
may contribute to it via contact diagrams in which an internal line
associated with either a $Z$ boson or a photon has been amputated, for
instance when the equations of motion are used to replace the term
$\partial_\mu B^{\mu \nu}$ with the respective current. Therefore, the
irreducible operators deserve a more careful study than the reducible
ones. We will present thus the Lagrangians and vertex functions in the
irreducible case, whereas in the reducible case we will list only the
respective operators and the Lagrangian prescribing the off-shell
electromagnetic properties of the $Z$ boson. In the next section we will
enumerate the operators of dimension eight that generate TNGBC.

\subsection{Irreducible operators}

We begin by classifying those operators which cannot be eliminated using
the equations of motion. We will categorize them according to $CP$
symmetry.

\subsubsection{$CP$-odd operators}

The operators we are interested in have the form ${\cal O}_i \partial^\rho
{\cal O}_j$, where ${\cal O}_i$ is any of the $ \mathrm{SU(2)_{L} \times
U(1)_{Y}} $ invariant expressions shown in (\ref{builbloc}). Given these
operators it is immediate to construct the new ones $(\partial^\rho {\cal
O}_i) {\cal O}_j$, which also belong to the irreducible group, but they
are not independent at all since they are related to the original
operators through a surface term. Bearing this in mind, we obtain the
following four independent $CP$-odd operators of dimension eight

\begin{eqnarray}
{\cal O}_{WW1}&=&i2\partial^\lambda(\Phi^\dag D_\mu \Phi){\rm
Tr}\left[{\bf W}^{\mu \nu}{\bf W}_{\lambda \nu}\right]+h.c.,\\ {\cal
O}_{WB1}&=&i(\Phi^\dag {\bf W}_{\mu \nu} D_\lambda \Phi)\partial^\lambda
B^{\mu \nu}+h.c., \\ {\cal O}_{WB2}&=&i(\Phi^\dag {\bf W}_{\mu \nu}
D_\lambda \Phi)\partial^\mu B^{\lambda \nu}+h.c., \\ {\cal
O}_{BB1}&=&i(\Phi^\dag D_\mu \Phi)B_{\lambda \nu}\partial^\lambda B^{\mu
\nu}+h.c. 
\end{eqnarray}

\noindent Notice that the operator ${\cal O}_{BB1}$ contains three $
\mathrm{SU(2)_{L} \times U(1)_{Y}} $ invariant structures which can be
contracted with the ordinary derivative in three different ways, leading
to the same number of operators. One of them, namely $i\partial^\lambda
(\Phi^\dag D_\mu \Phi)B_{\lambda \nu} B^{\mu \nu}$, is irreducible, but
can be expressed by means of integration by parts in terms of ${\cal
O}_{BB1}$ and the reducible operator $i(\Phi^\dag D_\mu
\Phi)(\partial^\lambda B_{\lambda \nu})B^{\mu \nu}$, which will be
considered later. 

\subsubsection{$CP$-odd structure of the $ZZZ$, $ZZ\gamma$, and $Z\gamma
\gamma$ couplings}

The 4-vector $Z_\mu$ arises from the term $(\Phi^\dag D_\mu \Phi)$ after
SSB, whereas the antisymmetric field tensors $F_{\mu \nu}$ and $Z_{\mu
\nu}$ appear through the relations

\begin{mathletters}
\begin{equation}
B_{\mu \nu}=c_w F_{\mu \nu}-s_w Z_{\mu \nu},
\end{equation}
\begin{equation}
W^3_{\mu \nu}=s_w F_{\mu \nu}+c_w Z_{\mu \nu}+ig(W^-_\mu W^+_\nu-W^+_\mu
W^-_\nu),
\end{equation}
\end{mathletters}

\noindent where $s_w(c_w)=sin\theta_w (cos\theta_w)$, with $\theta_w$ the
weak mixing angle. After the decomposition of these operators in terms of
the mass eigenstate fields, we are left with several Lorentz structures
corresponding to TNGBC, though not all of them are independent. Some of
them are identical, which is manifest after a subtle manipulation of their
Lorentz indices, whereas other ones are related through a surface term.
Consequently, the $ZZZ$, $ZZ\gamma$ and $Z\gamma \gamma$ couplings can be
described by the following independent Lorentz structures

\begin{equation}
{\cal L}^{CP-odd}_{L-ZZZ}=f^{ZZZ}_{L1}
Z_\lambda Z_{\mu \nu} \partial^\lambda Z^{\mu \nu}+f^{ZZZ}_{L2} Z_{\mu
\nu} Z^{\lambda \nu} \partial_\lambda Z^\mu,
\end{equation}

\begin{equation}
{\cal L}^{CP-odd}_{L-ZZ\gamma}=f^{ZZ\gamma}_{L1} Z^{\mu \nu}F_{\lambda
\nu}\partial^\lambda Z_\mu+f^{ZZ\gamma}_{L2} Z_\lambda Z_{\mu \nu}
\partial^\lambda F^{\mu \nu}+f^{ZZ\gamma}_{L3} Z_\lambda F_{\mu
\nu}\partial^\lambda Z^{\mu \nu},
\end{equation}

\begin{equation}
{\cal L}^{CP-odd}_{L-Z\gamma\gamma}=f^{Z\gamma\gamma}_{L1} F^{\mu
\nu}F_{\lambda \nu}\partial^\lambda Z_\mu+f^{Z\gamma\gamma}_{L2}
Z_\lambda F_{\mu \nu}\partial^\lambda F^{\mu \nu},
\end{equation}

\noindent where $L$ is a subscript standing for the linear scheme.
The coefficients $f^{ZZZ}_{Li}$ are defined by

\begin{mathletters}
\begin{equation}
f^{ZZZ}_{L1}=-\frac{s_{2w}}{4gm^2_Z}\Big[2(c_w\epsilon_{WB1}+s_w\epsilon_{BB1})+
c_w\epsilon_{WB2}\Big],\end{equation}
\begin{equation}
f^{ZZZ}_{L2}=-\frac{c^3_w}{gm^2_Z}\epsilon_{WW1},
\end{equation}
\end{mathletters}

\begin{mathletters}
\label{fZZgodd}
\begin{equation}
f^{ZZ\gamma}_{L1}=-\frac{c^3_w}{gm^2_Z}\epsilon_{WW1},
\end{equation}
\begin{equation}
f^{ZZ\gamma}_{L2}=\frac{c_w}{2gm^2_Z}\Big[2(c_w\epsilon_{WB1}+s_w\epsilon_{BB1})+
c_w\epsilon_{WB2}\Big],
\end{equation}
\begin{equation}
f^{ZZ\gamma}_{L3}=\frac{s_{2w}}{4gm^2_Z}\Big[2(c_w\epsilon_{BB1}-
s_w\epsilon_{WB1})- s_w\epsilon_{WB2}\Big],
\end{equation}
\end{mathletters}

\begin{mathletters}
\begin{equation}
f^{Z\gamma\gamma}_{L1}=-\frac{s_ws_{2w}}{2gm^2_Z}\epsilon_{WW1},
\end{equation}
\begin{equation}
f^{Z\gamma\gamma}_{L2}=-\frac{c^2_w}{2gm^2_Z}\Big[2(s_w\epsilon_{WB1}-
c_w\epsilon_{BB1})+ s_w\epsilon_{WB2}\Big],
\end{equation}
\end{mathletters}

\noindent with $s_{2w}=2s_wc_w$. We have also introduced the definition
$\epsilon_i=(m_Z/\Lambda)^4\alpha_i$, where $\alpha_i$ is the arbitrary
coefficient associated with each operator in the EL scheme. The vertex
functions are presented in Appendix I.

\subsubsection{$CP$-even operators}

Operators of this kind can be obtained from the $CP$-odd ones by replacing
each strength tensor with its respective dual, namely $\widetilde{{\bf
W}}_{\mu \nu}=(1/2)\epsilon_{\mu \nu \lambda \rho}{\bf W}^{\lambda \rho}$,
and a similar expression for $\widetilde{B}_{\mu \nu}$. There is a couple
of independent $CP$-even operators associated with each one of the
$CP$-odd operators ${\cal O}_{WW1}$, ${\cal O}_{WB2}$, and ${\cal
O}_{BB1}$. Note that in these operators both ${\bf W}$ tensors are
contracted via only one of their indices, leading to two independent
combinations of the dual tensor. On the other hand, in ${\cal O}_{WB1}$
the ${\bf W}$ and $B$ tensors appear contracted by both indices. Since the
two possible combinations of dual tensors are equivalent, just one
$CP$-even operator can be constructed from ${\cal O}_{WB1}$. In this way,
there are seven independent $CP$-even operators

\begin{eqnarray}
\label{eq11}
{\cal O}_{\widetilde{W}W1}&=&i2\partial^\lambda(\Phi^\dag D_\mu \Phi){\rm
Tr}\left[\widetilde{{\bf W}}^{\mu \nu}{\bf W}_{\lambda \nu}\right]+h.c.,\\
{\cal O}_{W\widetilde{W1}}&=&i2\partial^\lambda(\Phi^\dag D_\mu \Phi){\rm
Tr}\left[{\bf W}^{\mu \nu}\widetilde{{\bf W}}_{\lambda \nu}\right]+h.c.,\\
{\cal O}_{W\widetilde{B}1}&=&i(\Phi^\dag {\bf W}_{\mu \nu} D_\lambda
\Phi)\partial^\lambda \widetilde{B}^{\mu \nu}+h.c., \\ {\cal
O}_{\widetilde{W}B2}&=&i(\Phi^\dag \widetilde{{\bf W}}_{\mu \nu} D_\lambda
\Phi)\partial^\mu B^{\lambda \nu}+h.c., \\ {\cal
O}_{W\widetilde{B}2}&=&i(\Phi^\dag {\bf W}_{\mu \nu} D_\lambda
\Phi)\partial^\mu \widetilde{B}^{\lambda \nu}+h.c., \\ {\cal
O}_{\widetilde{B}B1}&=&i(\Phi^\dag D_\mu \Phi)\widetilde{B}_{\lambda
\nu}\partial^\lambda B^{\mu \nu}+h.c., \\ {\cal
O}_{B\widetilde{B}1}&=&i(\Phi^\dag D_\mu \Phi)B_{\lambda
\nu}\partial^\lambda \widetilde{B}^{\mu \nu}+h.c. 
\end{eqnarray}

\noindent We can make the ordinary derivative operate on the remaining $
\mathrm{SU(2)_{L} \times U(1)_{Y}} $ invariant terms out of which the
previous operators are constructed. The resulting operators are also of
the irreducible kind, but they are not independent since, as explained in
the $CP$-odd case, all of them are related to the first ones through a
surface term. 

\subsubsection{$CP$-even structure of the $ZZZ$, $ZZ\gamma$, and $Z\gamma
\gamma$ couplings}

After a careful analysis of the Lorentz structure induced by the $CP$-even
operators, we find that the $ZZZ$, $ZZ\gamma$, and $Z\gamma \gamma$
couplings are characterized, respectively, by two, five, and three
independent Lorentz structures

\begin{equation}
{\cal L}^{CP-even}_{L-ZZZ}=g^{ZZZ}_{L1} Z_\lambda Z_{\mu \nu}
\partial^\lambda \widetilde{Z}^{\mu \nu}+g^{ZZZ}_{L2} Z_\lambda Z_{\mu
\nu} \partial^\mu \widetilde{Z}^{\lambda \nu},
\end{equation}

\begin{eqnarray}
{\cal L}^{CP-even}_{L-ZZ\gamma}&=&g^{ZZ\gamma}_{L1}\widetilde{F}_{\mu
\nu}Z^{\lambda \nu}\partial_\lambda
Z^\mu+g^{ZZ\gamma}_{L2}\widetilde{Z}_{\mu \nu}F^{\lambda
\nu}\partial_\lambda Z^\mu +\nonumber \\
&&g^{ZZ\gamma}_{L3}\widetilde{Z}_{\lambda \nu}F^{\mu
\nu}\partial^\lambda Z_\mu+g^{ZZ\gamma}_{L4}Z^\lambda F^{\mu
\nu}\partial_\mu \widetilde{Z}_{\lambda \nu}+g^{ZZ\gamma}_{L5}Z^\lambda
Z^{\mu \nu}\partial_\mu \widetilde{F}_{\lambda \nu},
\end{eqnarray}

\begin{equation}
{\cal L}^{CP-even}_{L-Z\gamma \gamma}=g^{Z\gamma
\gamma}_{L1}\partial_\lambda Z^\mu \widetilde{F}_{\mu \nu}F^{\lambda
\nu}+g^{Z\gamma \gamma}_{L2}\partial_\lambda Z^\mu
\widetilde{F}^{\lambda \nu}F_{\mu \nu}+g^{Z\gamma \gamma}_{L3}Z_\lambda
F_{\mu \nu}\partial^\mu \widetilde{F}^{\lambda \nu},
\end{equation}

\noindent where the coefficients are

\begin{mathletters}
\begin{equation}
g^{ZZZ}_{L1}=\frac{s_{2w}}{4gm^2_Z}\Big[\frac{2
c^2_w}{s_w}(\epsilon_{\widetilde{W}W1}+ \epsilon_{W\widetilde{W}1})+2(c_w
\epsilon_{W\widetilde{B}1}-s_w \epsilon_{\widetilde{W}B1})+
c_w\epsilon_{\widetilde{W}B2}\Big],
\end{equation}
\begin{equation}
g^{ZZZ}_{L2}=\frac{s_{2w}}{4gm^2_Z}\Big[c_w \epsilon_{W\widetilde{B}2}-
2s_w \epsilon_{B\widetilde{B}1}\Big],
\end{equation}
\end{mathletters}

\begin{mathletters}
\label{gZZgeven}
\begin{equation}
\label{cZZgeven}
g^{ZZ\gamma}_{L1}=-
\frac{s_{2w}}{4gm^2_Z}\Big[2c_w(\epsilon_{\widetilde{W}W1}+
\epsilon_{\widetilde{B}B1})+s_w(\epsilon_{\widetilde{W}B2}+2\epsilon_{
W\widetilde{B}1})\Big],
\end{equation}
\begin{equation}
g^{ZZ\gamma}_{L2}=-\frac{s_{2w}}{4gm^2_Z}\epsilon_{\widetilde{W}W1},
\end{equation}
\begin{equation}
g^{ZZ\gamma}_{L3}=-
\frac{c^2_w}{2gm^2_Z}\Big[c_w(2\epsilon_{W\widetilde{B}1}+
\epsilon_{\widetilde{W}B2})+2s_w(\epsilon_{W\widetilde{W}1}+\epsilon_{
\widetilde{B}B1})\Big],
\end{equation}
\begin{equation}
g^{ZZ\gamma}_{L4}=\frac{c_w}{2gm^2_Z}\Big[s_w\epsilon_{W\widetilde{B}2}+
2c_w\epsilon_{B\widetilde{B}1}\Big],
\end{equation}
\begin{equation}
g^{ZZ\gamma}_{L5}=\frac{c^2_w}{2gm^2_Z}\Big[2s_w\epsilon_{B\widetilde{B}1}-
c_w\epsilon_{W\widetilde{B}2}\Big],
\end{equation}
\end{mathletters}

\begin{mathletters}
\begin{equation}
g^{Z\gamma \gamma}_{L1}=-\frac{c_ws_{2w}}{2gm^2_Z}\epsilon_{\widetilde{W}W1},
\end{equation}
\begin{equation}
g^{Z\gamma
\gamma}_{L2}=\frac{s_{2w}}{4gm^2_Z}\Big[c_w\epsilon_{\widetilde{W}B2}+
2(c_w\epsilon_{W\widetilde{B}1}-s_w\epsilon_{W\widetilde{W}1})+
2\frac{c^2_w}{s_w}\epsilon_{\widetilde{B}B1}\Big],
\end{equation}
\begin{equation}
g^{Z\gamma
\gamma}_{L3}=-\frac{c^2_w}{2gm^2_Z}\Big[s_w\epsilon_{W\widetilde{B}2}+
2c_w\epsilon_{B\widetilde{B}1}\Big]. 
\end{equation}
\end{mathletters}

\noindent The vertex functions are also presented in Appendix I. 

\subsection{Reducible operators}

The operators belonging to the reducible class are proportional to the
$\mathrm{SU(2)_{L} \times U(1)_{Y}} $ invariants $\partial^\mu (\Phi^\dag
D_\mu \Phi)$ and $\partial^\mu B_{\mu \nu}$. While those operators
containing the term $\partial^\mu (\Phi^\dag D_\mu \Phi)$ are
proportional to the scalar part of the $Z$ boson, those proportional to
the $\partial^\mu B_{\mu \nu}$ have the peculiarity that they generate the
Lorentz structures required to define the off-shell electromagnetic
properties of the $Z$ boson, namely the transition magnetic (electric)
dipole and quadrupole moments. All of these operators can be reduced to
others by using the equations of motion. To define these structures, it
will be necessary to include some operators of dimension ten, but as they
can always be expressed in terms of other operators we will content
ourselves with list them. We will also present the Lagrangian prescribing
the off-shell electromagnetic properties of the $Z$ boson. The operators
will be classified according to these properties. 

\subsubsection{Operators that generate the off-shell electromagnetic
properties of the $Z$ boson}

All these operators are proportional to the $ \mathrm{SU(2)_{L} \times
U(1)_{Y}} $ invariant $\partial^\mu B_{\mu \nu}$. There are four operators
of this class: one couple of $CP$-odd ones and another couple of $CP$-even
ones

\begin{eqnarray}
{\cal O}_{WB3}&=&i(\Phi^\dag {\bf W}^{\mu \nu} D_\mu \Phi)\partial^\lambda
B_{\lambda \nu}+h.c., \\ {\cal O}_{BB3}&=&i(\Phi^\dag D_\mu \Phi)B^{\mu
\nu}\partial^\lambda B_{\lambda \nu}+h.c., \\ {\cal
O}_{\widetilde{W}B3}&=&i(\Phi^\dag \widetilde{{\bf W}}^{\mu \nu} D_\mu
\Phi)\partial^\lambda B_{\lambda \nu}+h.c., \\ {\cal
O}_{\widetilde{B}B3}&=&i(\Phi^\dag D_\mu \Phi)\widetilde{B}^{\mu
\nu}\partial^\lambda B_{\lambda \nu}+h.c. 
\end{eqnarray}

\noindent To define the off-shell electromagnetic properties of the $Z$
boson, it is necessary to include the following operators of dimension ten

\begin{eqnarray}
{\cal O}^{10}_{WB}&=&i(\Phi^\dag {\bf W}^{\mu \nu} D_\lambda
\Phi)\partial_\mu \partial^\lambda \partial^\rho B_{\rho \nu}+h.c., \\
{\cal O}^{10}_{BB}&=&i(\Phi^\dag D_\lambda \Phi)B^{\mu \nu}\partial_\mu
\partial^\lambda \partial^\rho B_{\rho \nu}+h.c.,\\ {\cal
O}^{10}_{\widetilde{W}B}&=&i(\Phi^\dag \widetilde{{\bf W}}^{\mu \nu}
D_\lambda \Phi)\partial_\mu \partial^\lambda \partial^\rho B_{\rho
\nu}+h.c., \\ {\cal O}^{10}_{\widetilde{B}B}&=&i(\Phi^\dag D_\lambda
\Phi)\widetilde{B}^{\mu \nu}\partial_\mu \partial^\lambda \partial^\rho
B_{\rho \nu}+h.c. 
\end{eqnarray}

\noindent We have excluded any redundant operator, as the ones related
through a surface term. The operator ${\cal O}_{DB}=\Phi^\dag (D_\mu
D_\nu+D_\nu D_\mu)\Phi \partial^\mu \partial_\lambda B^{\lambda \nu}$,
which does not contribute to the electromagnetic properties of the $Z$
boson, can be eliminated by using the equations of motion. The Lorentz
structures defining the off-shell electromagnetic properties of the $Z$
boson can be conveniently parametrized by the following Lagrangian

\begin{equation}
{\cal L}_{ZZ\gamma}=-e \left[(h^Z_1F^{\mu \nu}+h^Z_3\widetilde{F}^{\mu
\nu})Z_\mu \frac{\partial^\lambda Z_{\lambda \nu}}{m^2_Z}+ (h^Z_2F^{\mu
\nu}+h^Z_4\widetilde{F}^{\mu \nu})Z^\lambda \frac{\partial_\mu
\partial_\lambda \partial^\rho Z_{\rho \nu}}{m^4_Z}\right],
\end{equation}

\noindent where the coefficients are

\begin{mathletters}
\begin{equation}
h^Z_1=-\frac{s^2_{2w}}{8s_we^2}\epsilon_8,
\end{equation}
\begin{equation}
h^Z_3=-\frac{s^2_{2w}}{8s_we^2}\widetilde{\epsilon}_8,
\end{equation}
\begin{equation}
h^Z_2=-\frac{s_{2w}}{s_we^6}\epsilon_{10},
\end{equation}
\begin{equation}
h^Z_4=-\frac{s_{2w}}{s_we^6}\widetilde{\epsilon}_{10}.
\end{equation}
\end{mathletters}

\noindent The parameters $\epsilon_8$ and $\epsilon_{10}$ depend on the
coefficients of the operators of dimension eight and ten, and are given by

\begin{mathletters}
\begin{equation}
\epsilon_8=s_w\epsilon_{WB3}+2c_w\epsilon_{BB3},
\end{equation}
\begin{equation}
\widetilde{\epsilon}_8=
s_w\epsilon_{\widetilde{W}B3}+2c_w\epsilon_{\widetilde{B}B3},
\end{equation}
\begin{equation}
\epsilon_{10}=s_w\epsilon^{10}_{WB}+2c_w\epsilon^{10}_{BB},
\end{equation}
\begin{equation}
\widetilde{\epsilon}_{10}=
s_w\epsilon^{10}_{\widetilde{W}B}+2c_w\epsilon^{10}_{\widetilde{B}B},
\end{equation}
\end{mathletters}

\noindent where $\epsilon^{10}_i=(m_Z/\Lambda )^6 \alpha_i$ are the
coefficients of the dimension ten operators. The transition moments are
defined as

\begin{mathletters}
\begin{equation}
\mu_Z=-\frac{e}{\sqrt{2}m_Z}\frac{E^2_\gamma}{m^2_Z}\left(h^Z_1-h^Z_2\right),
\end{equation}
\begin{equation}
Q^e_Z=-\frac{2\sqrt{10}e}{m^2_Z}h^Z_1,
\end{equation}
\begin{equation}
d_Z=-\frac{e}{\sqrt{2}m_Z}\frac{E^2_\gamma}{m^2_Z}\left(h^Z_3-h^Z_4\right),
\end{equation}
\begin{equation}
Q^m_Z=-\frac{2\sqrt{10}e}{m^2_Z}h^Z_3,
\end{equation}
\end{mathletters}

\noindent where $\mu_Z$ ($d_Z$) is the off-shell magnetic (electric)
dipole moment and $Q^m_Z (Q^e_Z)$ is the magnetic (electric) quadrupole
moment of the $Z$ boson. 

\subsubsection{Operators proportional to the scalar part of the $Z$ boson}

These operators are characterized by the $ \mathrm{SU(2)_{L} \times
U(1)_{Y}} $ invariant $\partial_\mu (\Phi^\dag D^\mu \Phi)$. There are
three $CP$-odd operators of this type

\begin{eqnarray}
{\cal O}_{WW2}&=&i2\partial_\lambda (\Phi^\dag D^\lambda \Phi){\rm
Tr}\left[{\bf W}_{\mu \nu}{\bf W}^{\mu \nu}\right]+h.c, \\ {\cal
O}_{BB2}&=&i\partial_\lambda (\Phi^\dag D^\lambda \Phi)B_{\mu \nu}B^{\mu
\nu}+h.c.,\\ {\cal O}_{D \Phi}&=&i\Phi^\dag(D_\mu D_\nu+D_\nu D_\mu)\Phi
\partial^\mu (\Phi^\dag D^\nu \Phi)+h.c. 
\end{eqnarray}

\noindent The last operator generates only the $ZZZ$ coupling, which can
be expressed by integration by parts as a coupling proportional to the
scalar part of the $Z$ boson. As for $CP$-even operators, there are only a
pair of this kind

\begin{eqnarray}
{\cal O}_{W\widetilde{W}2}&=&i2\partial_\lambda (\Phi^\dag D^\lambda
\Phi){\rm Tr}\left[{\bf W}_{\mu \nu}\widetilde{{\bf W}}^{\mu
\nu}\right]+h.c, \\ {\cal O}_{B\widetilde{B}2}&=&i\partial_\lambda
(\Phi^\dag D^\lambda \Phi)B_{\mu \nu}\widetilde{B}^{\mu \nu}+h.c. 
\end{eqnarray}

\noindent We have ignored any operator which can be expressed as a linear
combination of those given above.

\section{The nondecoupling scenario}
\label{nlc}

We will proceed now to consider the possibility that new physics effects
do not decouple from low energy physics. In this situation, the relevant $
\mathrm{SU(2)_{L} \times U(1)_{Y}} $ invariant structures are the same as
in the linear case, with the Higgs doublet being replaced by the following
unitary matrix

\begin{equation}
U=\exp[\frac{2i\sigma^i \phi^i}{v}],
\end{equation}

\noindent where the $\phi^i$ scalars would become Goldstone bosons. The
covariant derivative in the nonlinear realization of the $
\mathrm{SU(2)_{L} \times U(1)_{Y}} $ group is defined as ${\bf D}_\mu
U=\partial_\mu U+ig{\bf W}_\mu U-ig'U{\bf B}_\mu $, where the Abelian
field is now defined as ${\bf B}_\mu=(\sigma^3/2) B_\mu$. The basic
structures out of which TNGBC can be constructed are the $
\mathrm{SU(2)_{L} \times U(1)_{Y}} $ invariants ${\rm Tr}\left[\sigma^3
U^\dag {\bf D}_\mu U\right]$, ${\rm Tr}\left[U^\dag ({\bf D}_\mu {\bf
D}_\nu+{\bf D}_\nu {\bf D}_\mu)U\right]$, and ${\rm Tr}\left[U^\dag {\bf
W}_{\mu \nu}{\bf D}_\lambda U\right]$, which in mass units have dimension
one, two, and three. Like their linear counterparts, these invariants are
essential to construct any TNGBC because they induce the Lorentz
structures $Z_\mu$, $Z_\mu Z_\nu$, and $Z_\lambda Z_{\mu \nu}(F_{\mu
\nu})$. Since these structures have a lower dimension than their analogous
in the linear case, in the nonlinear scenario not only it is possible to
construct dimension six operators inducing TNGBC, but it is also possible
a larger number of independent operators. As we will show below, there
exist some operators of dimension four which induce the $ZZZ$ coupling,
though not the $ZZ\gamma$ and $Z\gamma \gamma$ ones. Nevertheless, such
operators are proportional to the scalar part of the $Z$ boson and belong
to the reducible group. We will use the same criterion used in the linear
case to classify all of the independent operators. We will refrain from
any technical detail already explained in the linear case if it is not
relevant for the present discussion. 

\subsection{Irreducible operators}

These operators are proportional to the $ \mathrm{SU(2)_{L} \times
U(1)_{Y}} $ invariant structures ${\rm Tr}\left[\sigma^3 U^\dag {\bf
D}_\mu U\right]$ and ${\rm Tr}\left[U^\dag {\bf W}_{\mu \nu}{\bf
D}_\lambda U\right]$. We will classify them according to $CP$ symmetry. 

\subsubsection{$CP$-odd operators}

The dimension six operators resembling those of the linear scenario are the
following

\begin{eqnarray}
{\cal L}_{WW1}&=&2i\frac{\lambda_{WW1}}{\Lambda^2}\partial^\lambda {\rm
Tr}\left[\sigma^3 U^\dag {\bf D}_\mu U\right]{\rm Tr}\left[{\bf W}^{\mu
\nu}{\bf W}_{\lambda \nu}\right]+h.c.,\\ {\cal
L}_{WB1}&=&i\frac{\lambda_{WB1}}{\Lambda^2}{\rm Tr}\left[U^\dag {\bf
W}_{\mu \nu}{\bf D}_\lambda U\right]\partial^\lambda B^{\mu \nu}+h.c.,\\
{\cal L}_{WB2}&=&i\frac{\lambda_{WB2}}{\Lambda^2}{\rm Tr}\left[U^\dag {\bf
W}_{\mu \nu}{\bf D}_\lambda U\right]\partial^\mu B^{\lambda \nu}+h.c.,\\
{\cal L}_{BB1}&=&i\frac{\lambda_{BB1}}{\Lambda^2}{\rm Tr}\left[\sigma^3
U^\dag {\bf D}_\mu U\right]B_{\lambda \nu}\partial^\lambda B^{\mu
\nu}+h.c.,
\end{eqnarray}

\noindent where we are using the symbol $\Lambda$, introduced in the
linear case, to denote the new physics scale. As the structure ${\rm
Tr}\left[\sigma^3 U^\dag {\bf D}_\mu U\right]$ has dimension one, we can
construct three new independent operators of dimension six which have no
dimension eight counterpart in the linear realization. They are given by

\begin{eqnarray}
{\cal L}_{DD}&=&i\frac{\lambda_{DD}}{\Lambda^2}{\rm Tr}\left[\sigma^3
U^\dag {\bf D}_\mu U\right]\Box {\rm Tr}\left[\sigma^3 U^\dag {\bf D}_\nu
U\right]\partial^\mu {\rm Tr}\left[\sigma^3 U^\dag {\bf D}_\mu
U\right]+h.c., \\ {\cal L}_{DB1}&=&\frac{\lambda_{DB1}}{\Lambda^2}{\rm
Tr}\left[\sigma^3 U^\dag {\bf D}_\mu U\right]\partial_\lambda {\rm
Tr}\left[\sigma^3 U^\dag {\bf D}_\nu U\right]\partial^\lambda B^{\mu
\nu}+h.c.,\\ {\cal L}_{DB2}&=&\frac{\lambda_{DB2}}{\Lambda^2}{\rm
Tr}\left[\sigma^3 U^\dag {\bf D}_\mu U\right]\partial_\nu {\rm
Tr}\left[\sigma^3 U^\dag {\bf D}_\lambda U\right]\partial^\lambda B^{\mu
\nu}+h.c. 
\end{eqnarray}

\noindent Note that in the linear scheme the operator corresponding to
${\cal L}_{DD}$ have dimension twelve, whereas those related to ${\cal
L}_{DB1}$ and ${\cal L}_{DB2}$ are of dimension ten. These operators have
the peculiarity that they induce TNGBC exclusively, {\it i.e.} there are
no interactions containing a charged $W$ boson, which can be seen by
noting that the structure ${\rm Tr}\left[\sigma^3 U^\dag {\bf D}_\mu
U\right]$ is proportional to the $Z_\mu$ boson in the unitary gauge. While
the first one of these operators induces only the $ZZZ$ coupling, the
remaining ones generate both the $ZZZ$ and $ZZ\gamma$ couplings. There is
no $Z\gamma \gamma$ coupling arising from these kind of operators, which
implies that the Lorentz structure of it is the same in both the linear
and the nonlinear realizations of the electroweak group, at least at this
order. 

\subsubsection{$CP$-odd structure of the $ZZZ$, $ZZ\gamma$, and $Z\gamma
\gamma$ couplings.}

After decomposing the nonlinear $CP$-odd operators in terms of the
physical fields, we have found that the $ZZZ$ coupling can be described by
five independent Lorentz structures, and so is the $Z Z \gamma$ vertex. On
the other hand, the $Z \gamma \gamma$ coupling becomes changed, as
compared to its counterpart in the linear case, in its coefficients but
not in its Lorentz structure. We thus have

\begin{equation}
{\cal L}^{CP-odd}_{NL-ZZZ}={\cal L}^{CP-odd}_{L-ZZZ}+f^{ZZZ}_{NL3}Z_\mu
\Box Z_\nu \partial^\mu Z^\nu+f^{ZZZ}_{NL4}Z_\mu \partial_\lambda Z_\nu
\partial^\lambda Z^{\mu \nu}+f^{ZZZ}_{NL5}Z_\mu \partial_\nu Z_\lambda
\partial^\lambda Z^{\mu \nu},
\end{equation}

\begin{equation}
{\cal L}^{CP-odd}_{NL-ZZ\gamma}={\cal
L}^{CP-odd}_{L-ZZ\gamma}+f^{ZZ\gamma}_{NL4}Z_\mu \partial_\lambda Z_\nu
\partial^\lambda F^{\mu \nu}+f^{ZZ\gamma}_{NL5}Z_\mu \partial_\nu
Z_\lambda \partial^\lambda F^{\mu \nu},
\end{equation}

\begin{equation}
{\cal L}^{CP-odd}_{NL-Z\gamma \gamma}={\cal L}^{CP-odd}_{L-Z\gamma \gamma},
\end{equation}

\noindent where the respective coefficients are obtained from those of the
linear scenario (Sec. \ref{lc}) through the relation

\begin{equation}
\label{relation}
f^{VVV}_{NLi}=\left(\frac{\Lambda}{m_Z}\right)^2f^{VVV}_{Li},
\end{equation}

\noindent with the remaining coefficients being given by

\begin{mathletters}
\begin{equation}
f^{ZZZ}_{NL3}=\frac{2g^3}{c^3_em^2_Z}\epsilon_{DD},
\end{equation}
\begin{equation}
f^{ZZZ}_{NL4}=-\frac{2g^2s_w}{c^2_wm^2_Z}\epsilon_{DB1},
\end{equation}
\begin{equation}
f^{ZZZ}_{NL5}=-\frac{2g^2s_w}{c^2_wm^2_Z}\epsilon_{DB2},
\end{equation}
\end{mathletters}

\begin{mathletters}
\begin{equation}
f^{ZZ\gamma}_{NL4}=\frac{2g^2}{c_wm^2_Z}\epsilon_{DB1},
\end{equation}
\begin{equation}
f^{ZZ\gamma}_{NL5}=\frac{2g^2}{c_wm^2_Z}\epsilon_{DB2}.
\end{equation}
\end{mathletters}

\noindent We have also introduced the definition
$\epsilon_i=(m_Z/\Lambda)^2\lambda_i$, where $\lambda_i$ represents the
coefficient associated with each operator of the nonlinear scenario. All
the vertex functions are given in Appendix II.

\subsubsection{CP-even operators}

There are eight operators belonging to this class. Seven of them can be
easily obtained from their linear counterparts whereas a new one is
obtained from the $CP$-odd operator ${\cal L}_{DB1}$ when the tensor
$B_{\mu \nu}$ is replaced by its dual. The $CP$-odd operator which is
equivalent to ${\cal L}_{DB2}$ is not independent as it generates TNGBC
with a Lorentz structure already induced by the operators resembling those
of the linear case. In this way, we are left with the following
independent $CP$-even operators

\begin{eqnarray}
{\cal \widetilde{L}}_{\widetilde{W}W1}&=
&2i\frac{\lambda_{\widetilde{W}W1}}{\Lambda^2}\partial^\lambda {\rm
Tr}\left[\sigma^3 U^\dag {\bf D}_\mu U\right]{\rm Tr}\left[\widetilde{{\bf
W}}^{\mu \nu}{\bf W}_{\lambda \nu}\right]+h.c.,\\ {\cal
\widetilde{L}}_{W\widetilde{W}1}&=
&2i\frac{\lambda_{W\widetilde{W}1}}{\Lambda^2}\partial^\lambda {\rm
Tr}\left[\sigma^3 U^\dag {\bf D}_\mu U\right]{\rm Tr}\left[{\bf W}^{\mu
\nu}\widetilde{{\bf W}}_{\lambda \nu}\right]+h.c.,\\ {\cal
\widetilde{L}}_{W\widetilde{B}1}&=
&i\frac{\lambda_{W\widetilde{B}1}}{\Lambda^2}{\rm Tr}\left[U^\dag {\bf
W}_{\mu \nu}{\bf D}_\lambda U\right]\partial^\lambda \widetilde{B}^{\mu
\nu}+h.c.,\\ {\cal \widetilde{L}}_{\widetilde{W}B2}&=
&i\frac{\lambda_{\widetilde{W}B2}}{\Lambda^2}{\rm Tr}\left[U^\dag
\widetilde{{\bf W}}_{\mu \nu}{\bf D}_\lambda U\right]\partial^\mu
B^{\lambda \nu}+h.c.,\\ {\cal \widetilde{L}}_{W\widetilde{B}2}&=
&i\frac{\lambda_{W\widetilde{B}2}}{\Lambda^2}{\rm Tr}\left[U^\dag {\bf
W}_{\mu \nu}{\bf D}_\lambda U\right]\partial^\mu \widetilde{B}^{\lambda
\nu}+h.c.,\\ {\cal \widetilde{L}}_{\widetilde{B}B1}&=
&i\frac{\lambda_{\widetilde{B}B1}}{\Lambda^2} {\rm Tr}\left[\sigma^3
U^\dag {\bf D}_\mu U\right]\widetilde{B}_{\lambda \nu}\partial^\lambda
B^{\mu \nu}+h.c.,\\ {\cal \widetilde{L}}_{B\widetilde{B}1}&=
&i\frac{\lambda_{B\widetilde{B}1}}{\Lambda^2} {\rm Tr}\left[\sigma^3
U^\dag {\bf D}_\mu U\right]B_{\lambda \nu}\partial^\lambda
\widetilde{B}^{\mu \nu}+h.c.,\\ {\cal \widetilde{L}}_{D\widetilde{B}1}&=
&i\frac{\lambda_{D\widetilde{B}1}}{\Lambda^2} {\rm Tr}\left[\sigma^3
U^\dag {\bf D}_\mu U\right]\partial_\lambda {\rm Tr}\left[\sigma^3 U^\dag
{\bf D}_\nu U\right]\partial^\lambda \widetilde{B}^{\mu \nu}+h.c. 
\end{eqnarray}

\subsubsection{$CP$-even structure of the $ZZZ$, $ZZ\gamma$, and $Z\gamma
\gamma$ couplings}

As far as their Lorentz structure is concerned, both the $ZZZ$ and the
$ZZ\gamma$ couplings differ from their analogues in the linear
realization. They receive a new contribution arising from the operator
${\cal \widetilde{L}}_{D\widetilde{B}1}$. The $ZZZ$ coupling in turn is
characterized by three independent Lorentz structures

\begin{equation}
{\cal L}^{CP-even}_{NL-ZZZ}={\cal
L}^{CP-even}_{L-ZZZ}+g^{ZZZ}_{NL3}Z_\mu \partial_\lambda Z_\nu
\partial^\lambda \widetilde{Z}^{\mu \nu}.
\end{equation}

\noindent The $ZZ\gamma$ coupling has six independent Lorentz structures
given by

\begin{equation}
{\cal L}^{CP-even}_{NL-ZZ\gamma}={\cal
L}^{CP-even}_{L-ZZ\gamma}+g^{ZZ\gamma}_{NL6}Z_\mu \partial_\lambda Z_\nu
\partial^\lambda \widetilde{F}^{\mu \nu}.
\end{equation}

\noindent The Lorentz structure of the $Z\gamma \gamma$ vertex coincides
with the one of its linear counterpart. As for the coefficients appearing
in the last equations, they are given in terms of the linear ones by means
of a relation similar to (\ref{relation}). The remaining coefficients are

\begin{equation}
g^{ZZZ}_{NL3}=\frac{2g^2s_w}{c^2_wm^2_Z}\epsilon_{D\widetilde{B}1},
\end{equation}

\begin{equation}
g^{ZZ\gamma}_{NL6}=-\frac{c_w }{s_w} g^{ZZZ}_{NL3}.
\end{equation}

\noindent The respective vertex function are presented in Appendix II.

\subsection{Reducible operators}

As was the case in the linear scenario, we can classify the reducible
operators in those contributing to the off-shell electromagnetic
properties of the $Z$ boson, and those which are proportional to the
scalar part of the $Z$ boson.

\subsubsection{Operators that generate the off-shell electromagnetic
properties of the $Z$ boson}

These operators are proportional to the $ \mathrm{SU(2)_{L} \times
U(1)_{Y}} $ invariant $\partial_\mu B^{\mu \nu}$, and are obtained from
their linear counterpart by replacing $\Phi^\dag D_\mu \Phi$ with ${\rm
Tr}\left[\sigma^3 U^\dag {\bf D}_\mu U\right]$. This give rise to
dimension six and dimension eight operators. The ones of dimension six are
given by

\begin{eqnarray}
{\cal L}_{WB3}&=&i\frac{\lambda_{WB3}}{\Lambda^2}{\rm Tr}\left[U^\dag {\bf
W}^{\mu \nu} {\bf D}_\mu U\right]\partial^\lambda B_{\lambda \nu}+h.c., \\
{\cal L}_{BB3}&=&i\frac{\lambda_{BB3}}{\Lambda^2}{\rm Tr}\left[U^\dag {\bf
D}_\mu U\right]B^{\mu \nu}\partial^\lambda B_{\lambda \nu}+h.c., \\ {\cal
L}_{\widetilde{W}B3}&= &i\frac{\lambda_{\widetilde{W}B3}}{\Lambda^2}{\rm
Tr}\left[U^\dag \widetilde{{\bf W}}^{\mu \nu} {\bf D}_\mu
U\right]\partial^\lambda B_{\lambda \nu}+h.c., \\ {\cal
L}_{\widetilde{B}B3}&= &i\frac{\lambda_{\widetilde{B}B3}}{\Lambda^2}{\rm
Tr}\left[\sigma^3 U^\dag {\bf D}_\mu U\right]\widetilde{B}^{\mu
\nu}\partial^\lambda B_{\lambda \nu}+h.c. 
\end{eqnarray}

\noindent Just as in the linear realization, there is another $CP$-odd
dimension six operator being given by

\begin{equation}
{\cal
L}_{DB}=\frac{\lambda_{DB}}{\Lambda^2}{\rm Tr}\left[U^\dag ({\bf D}_\mu {\bf D}_\nu+{\bf
D}_\nu {\bf D}_\mu)U\right]\partial^\mu \partial_\lambda B^{\lambda \nu},
\end{equation}

\noindent which, however, does not contributes to the electromagnetic
properties of the $Z$ boson. The operators of dimension eight, necessary
to an adequate definition of the electric and magnetic transition dipole
and quadrupole moments, are given by

\begin{eqnarray}
{\cal L}^8_{WB}&=&i\frac{\lambda^8_{WB}}{\Lambda^4}{\rm Tr}\left[U^\dag
{\bf W}^{\mu \nu} {\bf D}_\lambda U\right]\partial_\mu \partial^\lambda
\partial^\rho B_{\rho \nu}+h.c., \\ {\cal
L}^8_{BB}&=&i\frac{\lambda^8_{BB}}{\Lambda^4}{\rm Tr}\left[\sigma^3 U^\dag
{\bf D}_\lambda U\right]B^{\mu \nu}\partial_\mu \partial^\lambda
\partial^\rho B_{\rho \nu}+h.c.,\\ {\cal L}^8_{\widetilde{W}B}&=
&i\frac{\lambda^8_{\widetilde{W}B}}{\Lambda^4}{\rm Tr}\left[U^\dag
\widetilde{{\bf W}}^{\mu \nu} {\bf D}_\lambda U\right]\partial_\mu
\partial^\lambda \partial^\rho B_{\rho \nu}+h.c., \\ {\cal
L}^8_{\widetilde{B}B}&= &i\frac{\lambda^8_{\widetilde{B}B}}{\Lambda^4}{\rm
Tr}\left[\sigma^3 U^\dag {\bf D}_\lambda U\right]\widetilde{B}^{\mu
\nu}\partial_\mu \partial^\lambda \partial^\rho B_{\rho \nu}+h.c. 
\end{eqnarray}

\noindent They induce the off-shell electromagnetic properties of the $Z$
boson through the Lagrangian given in Sec. \ref{lc}. The coefficients
$h^Z_{1,3}$ and $h^Z_{2,4}$ are obtained from those of the linear scenario
after multiplying them by the factor $(\Lambda/m_Z)^2$ and
$(\Lambda/m_Z)^4$, respectively. 

\subsubsection{Operators that are proportional to the scalar part of the $Z$ boson}

These operators are proportional to the $ \mathrm{SU(2)_{L} \times
U(1)_{Y}} $ invariant $\partial_\mu {\rm Tr}\left[\sigma^3 U^\dag {\bf
D}^\mu U\right]$. As previously mentioned, there are a pair of dimension
four $CP$-odd operators which generate just the $ZZZ$ vertex. They are
given by

\begin{eqnarray}
{\cal L}^4_1&=&i\lambda_1 {\rm Tr}\left[\sigma^3 U^\dag {\bf D}_\nu
U\right]{\rm Tr}\left[\sigma^3 U^\dag {\bf D}^\nu U\right]\partial_\mu
{\rm Tr}\left[\sigma^3 U^\dag {\bf D}^\mu U\right]+h.c.,\\ {\cal
L}^4_2&=&i\lambda_2 {\rm Tr}\left[\sigma^3 U^\dag {\bf D}^\nu
U\right]\partial^\mu {\rm Tr}\left[U^\dag ({\bf D}_\mu {\bf D}_\nu+{\bf
D}_\nu {\bf D}_\mu)U\right] +h.c. 
\end{eqnarray}

The linear counterpart of the operator ${\cal L}^4_1$ has dimension ten,
while the one associated with ${\cal L}^4_2$ has dimension eight, as
described in Sec.  \ref{lc}. The remaining operators have dimension six,
and are obtained from those given in the linear case by the replacement of
$\Phi^\dag D_\mu \Phi$ by ${\rm Tr}\left[\sigma^3 U^\dag {\bf D}_\mu
U\right]$. There are four operators of this type: one pair of $CP$-odd
ones as well as one pair of $CP$-even ones

\begin{eqnarray}
{\cal L}_{WW2}&=&2i\frac{\lambda_{WW2}}{\Lambda^2}\partial_\lambda {\rm
Tr}\left[\sigma^3 U^\dag {\bf D}^\lambda U\right]{\rm Tr}\left[{\bf
W}_{\mu \nu}{\bf W}^{\mu \nu}\right]+h.c, \\ {\cal
L}_{BB2}&=&i\frac{\lambda_{BB2}}{\Lambda^2}\partial_\lambda {\rm
Tr}\left[U^\dag {\bf D}^\lambda U\right]B_{\mu \nu}B^{\mu \nu}+h.c.,\\
{\cal L}_{W\widetilde{W}2}&=
&2i\frac{\lambda_{\widetilde{W}W2}}{\Lambda^2}\partial_\lambda {\rm
Tr}\left[\sigma^3 U^\dag {\bf D}^\lambda U\right]{\rm Tr}\left[{\bf
W}_{\mu \nu}\widetilde{{\bf W}}^{\mu \nu}\right]+h.c, \\ {\cal
L}_{B\widetilde{B}2}&=
&i\frac{\lambda_{B\widetilde{B}2}}{\Lambda^2}\partial_\lambda {\rm
Tr}\left[U^\dag {\bf D}^\lambda U\right]B_{\mu \nu}\widetilde{B}^{\mu
\nu}+h.c. 
\end{eqnarray}

\section{Constraints from precision measurements}
\label{constraints}

Once a complete treatment of the effective operators inducing TNGBC has
been presented within both the linear and the nonlinear realizations of
the $ \mathrm{SU(2)_{L} \times U(1)_{Y}} $ gauge symmetry, our major
concern lies in how to get bounds on the respective coefficients of these
operators from current phenomenology. In this respect, considerable work
exists in the literature where bounds on anomalous trilinear gauge boson
couplings $W^+W^-\gamma$ have been analyzed. To this purpose, measurements
on some observables have been extensively used, such as the magnetic and
the electric dipole moments of elementary fermions, the $Z \to \bar{b} b$
branching fraction, as well as the processes $e^+e^- \to WW$ and $W^* \to
W \gamma$ \cite{ellison}. As for TNGBC, bounds on these couplings have
been obtained through the processes $e^+e^- \to Z \gamma (Z)$ and $q
\bar{q} \to Z \gamma (Z)$, although such studies involve only those
operators in which two gauge bosons are on-shell. To obtain bounds on our
operators, we will follow a similar approach as that in previous works. We
will also consider the rare decay $Z \to \nu \overline{\nu} \gamma$, which
is affected at tree level by TNGBC through the $ZZ\gamma$ vertex. Since
its SM contribution is insignificant \cite{hernandez}, this process might
offer an invaluable mode to unravel any latent new physics effect. 

\subsection{Decoupling scenario}

We will start by examining the situation in the decoupling scheme of EL.
Rather than performing any explicit calculation, it is worth to begin with
estimating on a general basis the expectations we should have as regards
to the size of TNGBC. It has been pointed out that persuasive theoretical
arguments indicate that one loop generated anomalous trilinear gauge boson
couplings are unlikely expected to be above the one percent level
\cite{burgess}. Indeed, the fact that TNGBC are induced at one loop level
suggests that they are of order $(g/4 \pi)^2$ in a wide class of models.
It has also been conjectured that even in theories with underlying strong
dynamics, trilinear gauge couplings are expected to have a sizeable
enhancement. In the SM, the $ZZ\gamma (Z)$ couplings are severely
constrained, even in the presence of a fourth fermion family they are
highly suppressed and thus out of the range of detectability
\cite{hernandez,r}. Regarding the bounds arising from phenomenological
grounds, we would like to begin by examining in a qualitative way whether
the current measurements on the magnetic and electric dipole moments of
elementary fermions can give any useful bound on TNGBC. 

\begin{figure}
\begin{center}
\epsfig{file=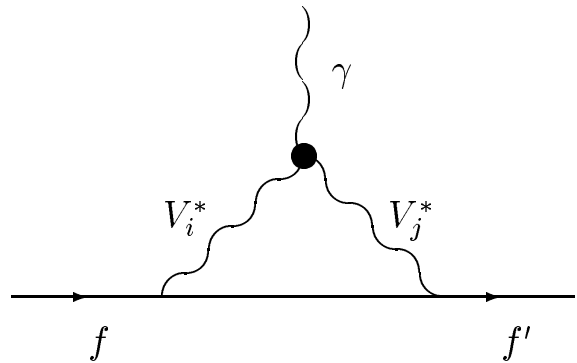,width=3in}
\caption{Contribution from TNGBC to the anomalous magnetic moment of fermions in
the effective Lagrangian approach.}
\label{fig:fig1}
\end{center}

\end{figure}

The effective operators presented so far not only induce TNGBC but also
anomalous $W^+W^-\gamma$ couplings. An exhaustive analysis on
phenomenological constraints would require to compute every contribution
to the observable under study, including also the ones coming from all of
the lower dimension operators inducing vertices which also affect the
process. For the sake of simplicity, a crude estimate can be obtained if
just some operators are considered at a time. In the specific case of the
magnetic moment of leptons, which receives contributions from $CP$-even
operators exclusively, a profuse work has been devoted to study
comprehensively the contributions from the lowest order effective
operators respecting the $ \mathrm{SU(2)_{L} \times U(1)_{Y}} $ gauge
invariance, linearly and nonlinearly realized, which induce nonstandard
anomalous couplings. In this respect, there are one loop generated
operators of dimension six which induce $W^+W^-\gamma$ couplings, but not
TNGBC. These operators contribute to the magnetic moment of leptons via
their insertion in the loop diagram depicted in Fig.\ \ref{fig:fig1}
\cite{martinez}. Secondly, some dimension six operators directly induce
the magnetic moment term at tree level, though they are generated at the
one loop level. Finally, the redefinition of the gauge fields, necessary
to an adequate definition of the quadratic part of the theory, also
affects the anomalous magnetic moment value. To obtain bounds on the
coefficients of the $CP$ conserving operators, the full contribution to
the anomalous magnetic moment of the muon was computed \cite{aew2}.  In
this respect, the strategy which has been found to be the most suitable
for estimating the size of loops involving an effective vertex is that of
dimensional regularization, together with the minimal subtraction
($\overline{\mathrm MS}$)  renormalization scheme. According to it and
retaining only the leading logarithmic dependence on the new physics scale
$\Lambda$, it was found that the contribution from dimension six operators
inducing the $W^+W^- \gamma$ vertex is given by

\begin{equation}
\label{estimate}
\delta a_\mu=\, \eta_0 \left(\frac{m_\mu}{\Lambda}\right)^2 \,
O(\log{\Lambda^2/m_W^2})  \, \alpha_{\mathrm{L}},
\end{equation}

\noindent where $\eta_0$ is a factor dependent on the particular graph,
and $\alpha_{\mathrm{L}}$ is directly related to the operator
coefficients. Numerically, one obtains from this equation $\left|\delta
a_\mu\right|/10^{-9}=\alpha_{\mathrm{L}} \left(1+\log \Lambda \right)
/\Lambda^2$.  If it is taken the accepted lowest value of 1 TeV for the
new physics scale $\Lambda$, we are left with the unpromising result that
the operator coefficient should be of order $O(1)$ to have any chance of
being detected. But this result is far beyond the estimate of $\alpha_L$
being of order $(g/4 \pi)^2$. Indeed, only the direct contribution is
expected to give a measurable contribution to the magnetic moment of the
muon. In view of this result, it is natural to think that we should not
expect a better situation for TNGBC since they are generated by higher
order operators. We note that dimension eight operators are suppressed by
the factor $(v/\Lambda)^2$, with $v=246$ GeV the vacuum expectation value,
with respect to dimension six operators. A rough estimate is obtained if
we multiply (\ref{estimate}) with the suppression factor and evaluate
again at $\Lambda=1$ TeV.  We obtain the discouraging result that
$\alpha_L$ should be of order $O(100)$, which is very unlikely to occur,
to allow any TNGBC to be experimentally detected. By way of illustration,
we have explicitly computed the contribution to the muon anomalous
magnetic moment which is obtained by introducing in the one loop diagram
of Fig.\ \ref{fig:fig1} the effective $ZZ \gamma$ vertex whose Feynman
rule is given by equations (\ref{cZZgeven}) and (\ref{feynrul}). After
isolating the divergent part, the application of the $\overline{\mathrm
MS}$ scheme gives

\begin{equation}
\label{amu}
\delta a_{\mu}^{L1}=\frac{t_w (4 s_w^2-1)g}{256 \pi^2}
\Big(\frac{m_z}{\Lambda}\Big)^2
\Big(\frac{m_\mu}{\Lambda}\Big)^2\left(\log\left(\frac{\Lambda}{m_Z}\right)^2+\frac{3}{4}\right)\widetilde{\epsilon}_{L1},
\end{equation}

\noindent where $\widetilde{\epsilon}_{L1}$ is the factor multiplying
$s_{2w}/(g m_Z^2)$ in the coefficient $g^{ZZ\gamma}_{L1}$ which is defined
in (\ref{cZZgeven}). In fact, if this equation is numerically evaluated we
find that the actual bound is looser than the rough estimate. We thus see
that it seems there is few hopes that a reasonable bound on $CP$-even
TNGBC could be obtained from precision measurements on the magnetic moment
of the muon. Although we have examined the situation of only one vertex,
the same result is expected for the remaining ones. In fact, as shown in
the appendices, the Lorentz structure which parametrizes TNGBC does not
differ essentially for each case. The best result would be obtained if all
contributions add up coherently, though there is no evident reason to
expect it. 

A similar analysis can be done for the $CP$-odd operators which contribute
to the electric dipole moment of fermions. In this case a strong bound,
from precision measurements on the electric dipole moment of neutron,
exists on dimension six operators inducing anomalous $W^+W^-\gamma$
couplings \cite{queijeiro}. The respective operator coefficients are
constrained to lie below the $10^{-3}$ level.  Since our $CP$-odd
operators, which also induce anomalous $W^+W^-\gamma$ couplings, are of
dimension eight in the decoupling scenario we could not expect to get a
better bound for their coefficients. Once more, a rough estimate would be
obtained by dividing the bound on dimension six operators by the
suppression factor $(v/\Lambda)^2$.

\begin{figure}
\begin{center}
\epsfig{file=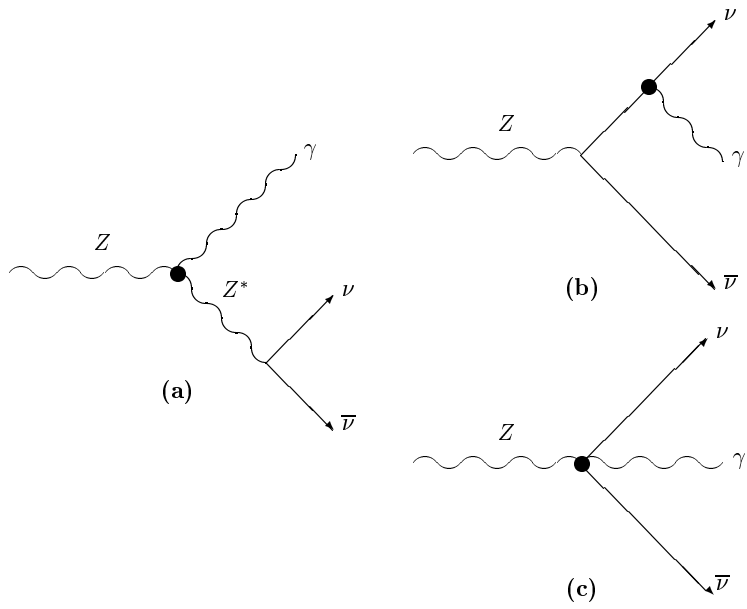,width=4in}
\caption{Feynman diagrams contributing to the decay $Z \to \nu \overline{\nu}
\gamma$ in the effective Lagrangian  approach.}
\label{fig:fig2}
\end{center}
\end{figure}

Now let us focus on the rare $Z$ boson decay $Z \to \nu \overline{\nu}
\gamma$, which has been studied within both the SM realm and the EL
approach \cite{maya,hernandez}. It was shown that the SM contribution
turns out to be negligible small, with a branching ratio of order
$10^{-10}$ \cite{hernandez}. In the EL approach, this process arises at
tree level, as depicted in Fig.\ \ref{fig:fig2}. In addition it has also
the advantage of receiving contributions from TNGBC only through the
$ZZ\gamma$ vertex, as depicted in Fig.\ \ref{fig:fig2} a.  Although there
are also lower dimension effective operators contributing to $Z \to \nu
\overline{\nu} \gamma$ through the Feynman diagrams of Fig.\
\ref{fig:fig2} b and \ref{fig:fig2} c \cite{maya}, we will not include
those contributions in here since they are not associated with TNGBC. 
Furthermore, we are only interested on estimating the best possible bound
on TNGBC.

The measurement of energetic single-photons at LEP arising from the decay
$Z \to \nu \overline{\nu} \gamma$ has been used to put a direct limit on
the magnetic moment of the $\tau$ neutrino \cite{Acc95}. For the purpose
of the present analysis, the search for the energetic single-photons
events on the data collected by the L3 collaboration may be translated
into bounds on TNGBC. In order to reduce backgrounds, the L3 collaboration
required the photon energy to be greater than one half the $e^+e^-$ beam
energy. It was obtained a limit on the branching ratio for $Z \to \nu
\overline{\nu} \gamma$ of one part in a million when the photon energy is
above 30 GeV \cite{Acc95}. To calculate the decay width, we will follow
closely the notation of \cite{hernandez}.  Expressing the invariant
amplitude $\cal M$ in terms of the variables $x=2k_1.p_1/m^2_Z$ and
$y=2k_1.p_2/m^2_Z$, the $Z(k_2)  \to A(k_1)\nu(p_1) \overline{\nu}(p_2)$
decay width is given by

\begin{equation}
\label{eqwidth}
\Gamma(Z \to \bar{\nu}\nu\gamma) =\frac{m_Z}{256 \pi^3}\int_{0}^{1}dx
\int_{0}^{1-x}dy\arrowvert \overline{\cal M}\arrowvert^2.
\end{equation}

\noindent We have not imposed any energy cutoff since we want to estimate
the TNGBC bounds in a conservative way. From the Feynman rules for the
$ZZ\gamma$ vertex in Appendix I is obtained

\begin{equation}
\label{mc}
\arrowvert \overline{\cal M}\arrowvert^2 =\frac{1}{32}
\Big(\left(x^2+y^2\right)\,\left(1 -x- y \right) - 4 x\,y\Big)
\Big(\alpha^2+\widetilde{\alpha}^2\Big),
\end{equation}

\begin{mathletters}
\begin{equation}
\label{cpodd}
\alpha \equiv \alpha_L=\alpha_{L1} +  \alpha_{L2} - \alpha_{L3},
\end{equation}
\begin{equation}
\label{cpeven}
\widetilde{\alpha} \equiv \widetilde{\alpha}_L=\widetilde{\alpha}_{L1} -
\widetilde{\alpha}_{L2} - \widetilde{\alpha}_{L3} + \widetilde{\alpha}_{L4} +
\widetilde{\alpha}_{L5}.
\end{equation}
\end{mathletters}

\noindent As natural, there is no interference between $CP$ violating and
$CP$ conserving couplings. The coefficients $\alpha_{Li}
\;(\widetilde{\alpha}_{Li})$, which in turn are related to the $CP$-odd
($CP$-even) operator coefficients, can be extracted from (\ref{fZZgodd})
and (\ref{gZZgeven}), being given by

\begin{mathletters}
\begin{equation}
\label{fzzgl}
\alpha_{Li} =\left(\frac{g \,m_Z^2}{c_w}\right)\,
f^{ZZ\gamma}_{Li},
\end{equation}
\begin{equation}
\label{gzzgl}
\widetilde{\alpha}_{Li}=\left(\frac{g\,m_Z^2
}{c_w}\right)g^{ZZ\gamma}_{Li}.
\end{equation}
\end{mathletters}

\noindent After performing the integration in equation
(\ref{eqwidth}) we have

\begin{equation}
\label{br}
BR(Z \to \bar{\nu}\nu\gamma) = 2.912 \times 10^{-5}
\Big(\alpha_{L}^2+\widetilde{\alpha}_{L}^2\Big).
\end{equation}

\noindent taking the value $\Lambda=1$ TeV and considering the L3 bound on
the respective branching fraction, we obtain again the result that the
size of the $ZZ\gamma$ coupling should be beyond any reasonable
expectation to become perceptible through the process $Z \to \nu
\overline{\nu} \gamma$. Stated in other words, we may not expect moderate
bounds from this process. The reason of such a discouraging result is the
natural suppression of dimension eight operators. Our viewpoint would be
more pessimistic if we consider that in this calculation only those
contributions arising from effective operators inducing the $ZZ\gamma$
coupling have been included. However, there is no compelling reason to
disregard any other new physics contributions, such as the ones coming
from the Feynman diagrams shown in the figures 2a and 2b \cite{maya}.  In
view of our results, it is conceivable to state that any TNGBC associated
with underlying physics respecting linearly the $ \mathrm{SU(2)_{L} \times
U(1)_{Y}} $ symmetry would not be measurable through the processes
investigated in this work.  However, we cannot discard the case in which
certain TNGBC is given by a sum of loops whose contributions add up
coherently to give a large value.

\subsection{Nondecoupling scenario}

We will turn to analyze the situation in the nonlinear scenario, where
TNGBC are generated by dimension six operators. Therefore, we might expect
a better situation than that in the decoupling scenario. We will see that
the discussion for the linear scenario can be easily translated to
comprise the nonlinear case. To begin with, we can see from the Feynman
rules in appendix II that in the nonlinear scenario the $CP$-even
$ZZ\gamma$ vertex is parametrized by one extra Lorentz structure in
addition to those parametrizing this vertex in the decoupling case. The
result given in (\ref{amu}) for the linear scenario can be directly used
if we consider the substitution rule applicable to the operator
coefficients, that is $g^{ZZ\gamma}_{NLi}=(\Lambda /m_Z)^2
g^{ZZ\gamma}_{Li}$. We then have that the leading term obtained by
including in the loop graph of Fig.\ \ref{fig:fig1} the $ZZ\gamma$ vertex
associated with $g^{ZZ\gamma}_{NL1}$ is

\begin{equation}
\label{amunl}
\delta a_{\mu}^{NL1}=\frac{t_w (4 s_w^2-1)g}{256 \pi^2}
\Big(\frac{m_\mu}{v}\Big)^2\left(\log\left(\frac{v}{m_Z}\right)^2
+\frac{3}{4}\right)\widetilde{\epsilon}_{NL1},
\end{equation}

\noindent where we have employed the conservative value $\Lambda \to v$.
Numerically one obtains $\delta a_{\mu}^{NL1}=-0.767 \,\times
10^{-9}\widetilde{\epsilon}_{NL1}$. On the other hand, the data collected
through the BNL E281 experiment together with the SM predictions put a
bound on any new physics contribution to $a_{\mu}$ of $-7.1 \times 10^{-9}
< \delta a_{\mu} < 22.1 \times 10^{-9}$ at 95 \%CL \cite{pdg}.  As a
consequence, probing $\delta a_{\mu}^{NL}$ at the $\pm 10^{-9}$ level
provides a sensitivity to $\widetilde{\epsilon}_{NL}$ of about $O(1)$ at
most, which translates into a loose bound for the operator coefficients
$\lambda_i$. This situation is not better than the result obtained in
\cite{aew2} for the dimension four operators inducing $W^+W^-\gamma$
couplings within the nonlinear scheme. Moreover, as there are other
sources of new physics which can affect the anomalous magnetic moment, it
is hard to think that any TNGBC could be competitive in this process, even
in the nonlinear scenario. 

Regarding the rare decay $Z \to \nu \overline{\nu} \gamma$, after the
inclusion of all the contributions arising from the $ZZ\gamma$ vertex we
have that (\ref{mc})  remains valid, though (\ref{cpodd}) and
(\ref{cpeven}) now read

\begin{mathletters}
\begin{equation}
\label{cpoddnl}
\alpha \equiv \alpha_{NL} = \alpha_{NL1} +  \alpha_{NL2} -
\alpha_{NL3}+2\,\alpha_{NL4}+ \alpha_{NL5},
\end{equation}
\begin{equation}
\label{cpevennl}
\widetilde{\alpha} \equiv \widetilde{\alpha}_{NL}=\widetilde{\alpha}_{NL1} -
\widetilde{\alpha}_{NL2} - \widetilde{\alpha}_{NL3} + \widetilde{\alpha}_{NL4} +
\widetilde{\alpha}_{NL5} +2\,\widetilde{\alpha}_{NL6}.
\end{equation}
\end{mathletters}

\noindent The new coefficients $\alpha_{NLi}$ and
$\widetilde{\alpha}_{NLi}$ are obtained, with the adequate subscript
substitutions, via the relations (\ref{fzzgl})  and (\ref{gzzgl}), which
also hold for the nonlinear scenario. The coefficients
$f^{ZZ\gamma}_{NLi}$ have been given in previous sections. We will only
concentrate in the $CP$-conserving term, which has been widely studied in
the literature.  Equation (\ref{br}) and the L3 limit for the respective
branching ratio give the bound $\arrowvert \widetilde{\alpha}_{NL}
\arrowvert < 1.8 \times 10^{-1}$ if $\alpha_{NL}=0$. This is a more
promising result than that previously found in the linear scenario. In
fact, there exist a direct relation between the bound just obtained within
the nonlinear scenario and the ones presented elsewhere under the
parametrization derived in \cite{hpzh}. It will be shown below that
$\alpha_{NL}=2 g^2 h^Z_{10}/(c_w s_w)$ and $\widetilde{\alpha}_{NL}=2 g^2
h^Z_{30}/(c_w s_w)$ correspond to the low energy limit of the form factors
$h_{i}^Z$ used extensively to study the $ZZ\gamma$ vertex in the case in
which one $Z$ boson and the photon are on-shell \cite{baur2}. Our bound
translates thus into

\begin{equation}
\label{bound}
\arrowvert h^Z_{30} \arrowvert< 0.38
\end{equation}

\noindent if $h^Z_{10}=0$, which agrees with previous bounds \cite{pdg}.
Of course, the same result applies to $h^Z_{10}$ when $h^Z_{30}=0$. In
this analysis, we have considered that the SM contribution to the rare
decay $Z \to \nu \overline{\nu} \gamma$ is negligible, what is a good
approximation since it was found that the branching ratio is of order
$10^{-10}$ \cite{hernandez}. We have also neglected the contributions
coming from the operators which give rise to the effective vertices shown
in Fig.\ \ref{fig:fig2} b and \ref{fig:fig2} c. This is the most optimum
scenario indeed. It is likely that any TNGBC may be screened by any other
sources of new physics arising from lower dimension operators. Therefore,
a more comprehensive analysis must be done to disentangle any new physics
contributing to the processes $e^+e^- \; (q q)\to ZZ\gamma$ \cite{baur2}.

\subsection{Connection with results derived within the $\mathrm{U(1)_{em}}$
formalism}

In the last subsections, the EL parametrization we elaborated earlier was
used to examine the impact of TNGBC on some loop induced processes. It was
also examined whether it is possible to obtain any reasonable bound from
the current limit on the branching ratio of the rare decay $Z \to \nu
\overline{\nu} \gamma$.  These vertices were studied for the first time
long ago, although only one particle was allowed to be off-shell
\cite{hpzh}. Following this approach, it has been customary to parametrize
any new physics effects inducing TNGBC by certain structures derived out
of $\mathrm{U(1)_{em}}$ gauge invariance, Lorentz covariance, as well as
Bose symmetry, what corresponds to the so called $\mathrm{U(1)_{em}}$
framework \cite{glr}. The coefficients of such Lorentz structures are
taken to be form factors which actually comprise all our ignorance on the
underlying dynamics inducing TNGBC. In general, these form factors depend
on the squared momenta of the participating particles.  Furthermore, as
this dependence is unknown since the form factors are determined by the up
to now unknown physics, it is necessary to make some assumptions to
describe their behavior. This scheme has probed to be useful to constrain
the low energy values of the form factors through $Z\gamma$ production in
$e^+e^-$ and $qq$ collisions at LEP, the Tevatron, and the future LHC
\cite{glr}. 

The latter formalism is to be contrasted with the EL method followed in
this work, which in turn is well suited for studying new physics effects
in a model independent way and no form factors nor extra assumptions on
the unknown physics are required, but all our ignorance of the new physics
lies in dimensionless (or dimensionful)  coefficients associated with each
effective operator, which in turn only depend on the new physics energy
scale. Another peculiarity of the EL formalism is that we are allowed to
know what operators the new physics comes from, in contrast to the form
factor scheme where we only know that the form factors themselves are
generated at a given order in the $\mathrm{U(1)_{em}}$ effective
Lagrangian. To establish a direct connection between these two different
formalisms is not an immediate nor an easy task. In a previous work
\cite{r} both approaches, within the $\mathrm{U(1)_{em}}$ gauge invariant
scheme, were considered and their relation was established. It was shown
how the form factors are related to the coefficients associated with the
effective operators arising from the $\mathrm{U(1)_{em}}$ framework. At
this point, it is natural to ask whether there is a direct connection
between our own results, when it is considered the case of only one
off-shell particle, and those derived from the form factor
parametrization. We will show that in the case where the form factors are
given their low energy values $h_{i0}^Z$, there is a simple connection
indeed. 

To show the relation between our results and previous ones, we will
consider only the $ZZ\gamma$ coupling, in the specific case where both the
initial $Z$ boson and the photon are on-shell since it is the only
coupling involved in the rare process $Z \to \nu \overline{\nu} \gamma$.
The most general structure for the $ZZ\gamma$ vertex respecting Lorentz
covariance, $\mathrm{U(1)_{em}}$ gauge invariance and Bose symmetry is
given by

\begin{eqnarray} 
\label{zzgpar} 
\Gamma_{\alpha_1 \,\alpha_2
\,\alpha}^{ZZ\gamma}(k_1,k_2,k)=\frac{ie \left(k_2^2-m_Z^2 \right)}{m_Z^2}
&& \left( h_1^Z \left(k^{\alpha_1} g^{\alpha \alpha_2}-k^{\alpha_2}
g^{\alpha \alpha_1} \right)
+\frac{h_2^Z}{m_Z^2}k_2^{\alpha_1}\left(k_2\cdot k \, g^ {\alpha_2 \alpha}
-k^ {\alpha_2} k_2^{\alpha}\right) + \right. \\ \nonumber && \left. h_3^Z
\epsilon^{\alpha_1 \alpha_2 \alpha \mu} k^{\mu}+
\frac{h_4^Z}{m_Z^2}k_2^{\alpha_1}\epsilon^{\alpha_2 \alpha \mu \nu} k_{1
\mu} k_{2 \nu} \right)  
\end{eqnarray}

\noindent where all the momenta are taken as incoming. Any term
proportional to $k^{\alpha}$ and $k_1^{\alpha_1}$ has been omitted, the
same is true for those proportional to $k_2^{\alpha_2}$ because it is also
assumed that the virtual $Z$ boson couples to light fermions, as actually
happens in the decay $Z \to \ell \ell \gamma$. In this parametrization,
the $CP$-conserving terms $h_{1,2}^Z$ as well as the $CP$-violating ones
$h_{3,4}^Z$ are taken as form factors which depend on the dynamics of the
underlying new physics, in general they are unknown functions of the
squared momenta of the neutral bosons, namely $k^2$, $k_1^2$, and $k_2^2$.
Within the $\mathrm{U(1)_{em}}$ formalism, as far as the form factors
$h_{1,3}^Z$ are concerned, they receive contributions from dimension six
operators, whereas the ones $h_{2,4}^Z$ can be induced by dimension eight
or higher operators. Based on unitarity requirement, some authors have
extensively used the approximation $h_i^Z=h_{i0}^Z/(1+s/\Lambda^2)^n$,
with $n$ an integer, $h_{i0}^Z$ the form factor low energy value, and $s$
the squared momentum of the virtual $Z$ boson \cite{baur2}. If the energy
scale $\Lambda$ associated with the new physics inducing TNGBC is larger
than the energy scale involved in the process, {\it i.e.} the squared
momentum of the virtual particle, it is a good approximation to use the
low energy values of the form factors.  After the replacement $h_i^Z \to
h_{i0}^Z$ is done in (\ref{zzgpar}), we are left with the expression for
the $ZZ\gamma$ vertex which must coincide with the one obtained from our
results in the nonlinear scenario. 

Considering the assumptions just described, we can obtain from the
appendices the expression for the $ZZ\gamma$ vertex arising from the lower
dimension operators within either the linear scenario or the nonlinear
one. The $CP$-odd part is given by

\begin{equation}
\Gamma^{ZZ\gamma}_{\alpha_1 \alpha_2 \alpha}(k_1,k_2,k)=g_1 \left(k_2^2 -
m_Z^2\right)\epsilon_{\alpha_1 \alpha_2 \alpha\mu}k^\mu+g_2
\Big[k_{\alpha_1}\epsilon_{\alpha \alpha_2 \mu
\nu}k^\mu_1k^\nu_2-k_{\alpha_2}\epsilon_{\alpha \alpha_1 \mu
\nu}k^\mu_1k^\nu_2\Big],
\end{equation}

\noindent while the $CP$-even part is

\begin{eqnarray}
\widetilde{\Gamma}^{ZZ\gamma}_{\alpha_1 \alpha_2 \alpha}(k_1,k_2,k)&=&f_1
\left(k_2^2 - m_Z^2\right) \left(k_{\alpha_2}\,g_{\alpha_1
\alpha}-k_{\alpha_1}\,g_{\alpha_2 \alpha}\right) \\ \nonumber &+& f_2
\Big[k_{\alpha_1}\left(k_{1\alpha}k_{\alpha_2}-k_1\cdot k\, g_{\alpha_2
\alpha}\right)+k_{\alpha_2}\left(k_{2\alpha}k_{\alpha_1}-k_2\cdot k\,
g_{\alpha_1 \alpha}\right)\Big],
\end{eqnarray}

\noindent where the factors $f_i$ and $g_i$ are related to the
coefficients $f_i^{ZZ\gamma}$ and $g_i^{ZZ\gamma}$, respectively. At first
sight it seems there is no direct coincidence of the terms which multiply
$f_2$ and $g_2$ with those multiplying $h_{1}^Z$ and $h_{3}^Z$ in
(\ref{zzgpar}). However, after a judicious manipulation and with the aid
of Shouten's identity, it can be shown that this is the case indeed. We
thus obtain a simple expression for the $ZZ\gamma$ vertex

\begin{equation}
\Gamma^{ZZ\gamma}_{\alpha_1 \alpha_2 \alpha}(k_1,k_2,k)=\left(k_2^2 -
m_Z^2\right)\Big[g^{ZZ\gamma} \, \left(k_{\alpha_2}\,g_{\alpha_1
\alpha}-k_{\alpha_1}\,g_{\alpha_2 \alpha}\right)+f^{ZZ\gamma} \,
\epsilon_{\alpha_1 \alpha_2 \alpha\mu}k^\mu \Big],
\end{equation}

\noindent which do show an obvious connection with (\ref{zzgpar}). Instead
of giving explicit expressions for $f^{ZZ\gamma}$ and $g^{ZZ\gamma}$, we
will establish the connection of $h_{10}^Z$ and $h_{30}^Z$ with the
coefficients $\alpha$ and $\widetilde{\alpha}$ appearing in
(\ref{cpodd})-(\ref{cpeven}) for the linear scenario and
(\ref{cpoddnl})-(\ref{cpevennl}) for the nonlinear case. We thus have

\begin{mathletters}
\begin{equation}
h^Z_{10}=\frac{c_w \,s_w\, \alpha_L}{2 g^2 },
\end{equation}
\begin{equation}
h^Z_{30}=\frac{c_w \,s_w\, \widetilde{\alpha}_L}{2 g^2},
\end{equation}
\end{mathletters}

\noindent the same relation holds for these coefficients in the nonlinear
scenario.  Explicit expressions from $f^{ZZ\gamma}$ and $g^{ZZ\gamma}$ can
be easily extracted from (\ref{fzzgl}) and (\ref{gzzgl}). 

Finally, a few comments are in order. Although the $ZZ\gamma$ has the same
Lorentz structure in both realizations of the $ \mathrm{SU(2)_{L} \times
U(1)_{Y}} $ gauge symmetry, the main difference is that the operators
inducing these structures are of dimension six in the nonlinear scenario,
whereas in the linear case they are induced by dimension eight operators.
As a result, though the bounds found for the coefficients $h_{10}^Z$ and
$h_{30}^Z$ apply in both scenarios, if they were translated into the
operator coefficients $\alpha_i$ and $\lambda_i$, looser bounds would be
obtained in the linear scenario. Regarding the remaining TGBC, a similar
analysis following the lines sketched above was done for the $ZZZ$ and
$Z\gamma\gamma$ couplings. It was found that our results agree with those
previously presented.  Another interesting point to be noted is that,
since the operators which induce the most general TNGBC vertices also
induce those couplings with only one off-shell particle, any bound which
has been put on the latter will be immediately applicable to the former.

\section{Conclusions}
\label{conclu}

In this work we have presented an analysis of trilinear neutral gauge
boson couplings, $ZZZ$, $ZZ\gamma$ and $Z\gamma \gamma$, under the context
of the effective Lagrangian approach, both in the linear and the nonlinear
realizations of the $ \mathrm{SU(2)_{L} \times U(1)_{Y}} $ gauge symmetry. 
Particular emphasis has been given to the linear scenario since current
literature lacks of an analysis in this line. The most general case with
three off-shell bosons is considered. In the linear scenario these
couplings receive contributions from dimension eight operators, whereas in
the nonlinear scenario they are induced by dimension six operators. For
completeness, we have included the Lorentz structure which parametrizes
these vertices, ready to be used in any future calculation. Based on
general considerations and actual calculations, we conclude that, if the
until now unknown physics underlying the SM is of a decoupling nature, it
is not expected that TNGBC could have a considerable impact either through
their virtual effects or via direct production. In contrast, if new
physics effects arise from a strong coupling regime at higher energies
which is responsible for the breaking of the $ \mathrm{SU(2)_{L} \times
U(1)_{Y}} $ symmetry (endowing the gauge bosons with mass), the
possibility of measuring their effects still remains. The EL approach
indicates that, owing to the suppression of the operators inducing TNGBC,
it is difficult that the effects arising from them may compete with those
coming from other sources of new physics induced by lower dimension
operators. However, it may happen that some fortuitous fact, such as some
resonant effect, could give rise to large TNGBC in a particular model. In
this context, it would be useful a study in a model dependent way to have
more evidences which could lead us to a deeper understanding of TNGBC.

\section*{Acknowledgments}

Financial support from CONACYT and SNI (M\'exico) is acknowledged.

\section*{Appendix I: TNGBC Vertex functions in the linear scenario}
\label{a1}

In this appendix the vertex functions for the $ZZZ$, $ZZ\gamma$, and
$Z\gamma \gamma$ couplings within the linear scenario are presented. We
consider the most general case where the three bosons are virtual. The
particle momenta are denoted as described below and will be taken as
incoming everywhere. The $CP$-odd vertex functions are given by\\

$Z_{\alpha_1}(k_1)Z_{\alpha_2}(k_2)Z_{\alpha_3}(k_3)$ vertex:

\begin{mathletters}
\begin{equation}
\Gamma^{L-ZZZ}_{\alpha_1 \alpha_2
\alpha_3}(k_1,k_2,k_3)=\sum^2_{i=1}f^{ZZZ}_{Li}
\Gamma^{Li-ZZZ}_{\alpha_1 \alpha_2 \alpha_3}(k_1,k_2,k_3),
\end{equation}
\begin{eqnarray}
\Gamma^{L1-ZZZ}_{\alpha_1 \alpha_2
\alpha_3}(k_1,k_2,k_3)&=&2k_{1\alpha_1}(k_{2\alpha_3}k_{3\alpha_2}-
k_2\cdot k_3 g_{\alpha_2 \alpha_3})+\nonumber \\
&&2k_{2\alpha_2}(k_{1\alpha_3}k_{3\alpha_1}-k_1\cdot k_3 g_{\alpha_3
\alpha_3}) +2k_{3\alpha_3}(k_{1\alpha_2}k_{2\alpha_1}-k_1\cdot k_2
g_{\alpha_1 \alpha_2}), \\ \Gamma^{L2-ZZZ}_{\alpha_1 \alpha_2
\alpha_3}(k_1,k_2,k_3)&=&k_{1\alpha_2}(k_2\cdot k_3 g_{\alpha_1
\alpha_3}-k_{3\alpha_1}k_{2\alpha_3})+k_{1\alpha_3}(k_2\cdot k_3
g_{\alpha_1 \alpha_2}-k_{2\alpha_1}k_{3\alpha_2})\nonumber \\
&+&k_{2\alpha_1}(k_1\cdot k_3 g_{\alpha_2
\alpha_3}-k_{1\alpha_3}k_{3\alpha_2})+k_{2\alpha_3}(k_1\cdot k_3
g_{\alpha_1 \alpha_2}-k_{3\alpha_1}k_{1\alpha_2})\nonumber \\
&+&k_{3\alpha_1}(k_1\cdot k_2 g_{\alpha_2
\alpha_3}-k_{1\alpha_2}k_{2\alpha_3})+k_{3\alpha_2}(k_1\cdot k_2
g_{\alpha_2 \alpha_3}-k_{1\alpha_3}k_{2\alpha_1}).
\end{eqnarray}
\end{mathletters}

$Z_{\alpha_1}(k_1)Z_{\alpha_2}(k_2)A_\alpha(k)$ vertex:

\begin{mathletters}
\begin{equation}
\Gamma^{L-ZZ\gamma}_{\alpha_1 \alpha_2
\alpha}(k_1,k_2,k)=\sum^3_{i=1}f^{ZZ\gamma}_{Li}
\Gamma^{Li-ZZ\gamma}_{\alpha_1 \alpha_2 \alpha}(k_1,k_2,k),
\end{equation}
\begin{eqnarray}
\Gamma^{L1-ZZ\gamma}_{\alpha_1 \alpha_2
\alpha}(k_1,k_2,k)&=&k_{1\alpha_2}(k_2\cdot k g_{\alpha_1
\alpha}-k_{2\alpha}k_{\alpha_1})+k_{2\alpha_1}(k_1\cdot k g_{\alpha_2
\alpha}-k_{1\alpha}k_{\alpha_2}), \\ \Gamma^{L2-ZZ\gamma}_{\alpha_1
\alpha_2 \alpha}(k_1,k_2,k)&=&k_{\alpha_1}(k_2\cdot k g_{\alpha_2
\alpha}-k_{2\alpha}k_{\alpha_2})+k_{\alpha_2}(k_1\cdot k g_{\alpha_1
\alpha}-k_{1\alpha}k_{\alpha_1}), \\ \Gamma^{L3-ZZ\gamma}_{\alpha_1
\alpha_2 \alpha}(k_1,k_2,k)&=&k_{1\alpha_2}(k_1\cdot k g_{\alpha_1
\alpha}-k_{1\alpha}k_{\alpha_1})+k_{2\alpha_1}(k_2\cdot k g_{\alpha_2
\alpha}-k_{2\alpha}k_{\alpha_2}).
\end{eqnarray}
\end{mathletters}

$A_{\alpha_1}(k_1)A_{\alpha_2}(k_2)Z_\alpha(k)$ vertex:

\begin{mathletters}
\begin{equation}
\Gamma^{L-Z\gamma \gamma}_{\alpha_1 \alpha_2
\alpha}(k_1,k_2,k)=\sum^2_{i=1}f^{Z\gamma \gamma}_{Li}
\Gamma^{Li-Z\gamma \gamma}_{\alpha_1 \alpha_2 \alpha}(k_1,k_2,k),
\end{equation}
\begin{eqnarray}
\Gamma^{L1-Z\gamma \gamma}_{\alpha_1 \alpha_2
\alpha}(k_1,k_2,k)&=&k_1\cdot k(k_{2\alpha}g_{\alpha_1
\alpha_2}-k_{2\alpha_1}g_{\alpha_2 \alpha})+k_2\cdot
k(k_{1\alpha}g_{\alpha_1 \alpha_2}-k_{1\alpha_2}g_{\alpha_1
\alpha})\nonumber \\ &&+k_{\alpha_1}(k_1\cdot k_2g_{\alpha_2
\alpha}-k_{1\alpha_2}k_{2\alpha})+ k_{\alpha_2}(k_1\cdot k_2g_{\alpha_1
\alpha}-k_{2\alpha_1}k_{1\alpha}), \\ \Gamma^{L2-Z\gamma
\gamma}_{\alpha_1 \alpha_2 \alpha}(k_1,k_2,k)&=&k_\alpha
(k_{1\alpha_2}k_{2\alpha_1}-k_1\cdot k_2 g_{\alpha_1 \alpha_2}).
\end{eqnarray}
\end{mathletters}

Following the same conventions, the $CP$-even vertex functions are given by\\

$ZZZ$ vertex:

\begin{mathletters}
\begin{equation}
\widetilde{\Gamma}^{L-ZZZ}_{\alpha_1 \alpha_2
\alpha_3}(k_1,k_2,k_3)=\sum^2_{i=1}g^{ZZZ}_{Li}
\widetilde{\Gamma}^{Li-ZZZ}_{\alpha_1 \alpha_2 \alpha_3}(k_1,k_2,k_3),
\end{equation}
\begin{eqnarray}
\widetilde{\Gamma}^{L1-ZZZ}_{\alpha_1 \alpha_2
\alpha_3}(k_1,k_2,k_3)&=&k_{1\alpha_1}\epsilon_{\alpha_2 \alpha_3 \mu
\nu}k^\mu_2 k^\nu_3+k_{2\alpha_2}\epsilon_{\alpha_1 \alpha_3 \mu
\nu}k^\mu_1 k^\nu_3+k_{3\alpha_3}\epsilon_{\alpha_1 \alpha_2 \mu \nu}
k^\mu_1 k^\nu_2,\\
\widetilde{\Gamma}^{L2-ZZZ}_{\alpha_1 \alpha_2
\alpha_3}(k_1,k_2,k_3)&=&-\widetilde{\Gamma}^{L1-
ZZZ}_{\alpha_1 \alpha_2 \alpha_3}(k_1,k_2,k_3)+k_1\cdot k_2
\epsilon_{\alpha_1 \alpha_2 \alpha_3 \mu}(k_1-k_2)^\mu \nonumber \\
&+&k_1\cdot k_3 \epsilon_{\alpha_1 \alpha_2 \alpha_3
\mu}(k_3-k_1)^\mu+k_2\cdot k_3 \epsilon_{\alpha_1 \alpha_2 \alpha_3
\mu}(k_2-k_3)^\mu.
\end{eqnarray}
\end{mathletters}

$ZZ\gamma$ vertex:

\begin{mathletters}
\begin{equation}
\widetilde{\Gamma}^{L-ZZ\gamma}_{\alpha_1 \alpha_2
\alpha_3}(k_1,k_2,k)=\sum_{i=1}^5
\widetilde{\Gamma}^{Li-ZZ\gamma}_{\alpha_1 \alpha_2 \alpha}(k_1,k_2,k),
\end{equation}
\begin{eqnarray}
\label{feynrul}
\widetilde{\Gamma}^{L1-ZZ\gamma}_{\alpha_1 \alpha_2
\alpha}(k_1,k_2,k)&=&k_{1\alpha_2}\epsilon_{\alpha \alpha_1 \mu
\nu}k^\mu_1k^\nu_2-k_{2\alpha_1}\epsilon_{\alpha \alpha_2 \mu
\nu}k^\mu_1k^\nu_2, \\
\widetilde{\Gamma}^{L2-ZZ\gamma}_{\alpha_1
\alpha_2 \alpha}(k_1,k_2,k)&=&k_1\cdot k\epsilon_{\alpha_1 \alpha_2
\alpha \mu}k^\mu_2-k_2\cdot k\epsilon_{\alpha_1 \alpha_2 \alpha
\mu}k^\mu_1-k_\alpha \epsilon_{\alpha_1 \alpha_2 \mu
\nu}k^\mu_1k^\nu_2,\\
\widetilde{\Gamma}^{L3-ZZ\gamma}_{\alpha_1
\alpha_2 \alpha}(k_1,k_2,k)&=&k_{\alpha_2}\epsilon_{\alpha \alpha_1 \mu
\nu}k^\mu_1k^\nu_2-k_{\alpha_1}\epsilon_{\alpha \alpha_2 \mu
\nu}k^\mu_1k^\nu_2, \\
\widetilde{\Gamma}^{L4-ZZ\gamma}_{\alpha_1
\alpha_2 \alpha}(k_1,k_2,k)&=&k_2\cdot k\epsilon_{\alpha_1 \alpha_2
\alpha \mu}k^\mu_2-k_1\cdot k\epsilon_{\alpha_1 \alpha_2 \alpha
\mu}k^\mu_1-k_\alpha \epsilon_{\alpha_1 \alpha_2 \mu
\nu}k^\mu_1k^\nu_2,\\
\widetilde{\Gamma}^{L5-ZZ\gamma}_{\alpha_1
\alpha_2 \alpha}(k_1,k_2,k)&=&k_{\alpha_2}\epsilon_{\alpha \alpha_1 \mu
\nu}k^\mu_1k^\nu_2-k_{\alpha_1}\epsilon_{\alpha \alpha_2 \mu
\nu}k^\mu_1k^\nu_2+k\cdot (k_1-k_2)\epsilon_{\alpha_1 \alpha_2 \alpha
\mu}k^\mu.
\end{eqnarray}
\end{mathletters}

$Z\gamma \gamma$ vertex:

\begin{mathletters}
\begin{equation}
\widetilde{\Gamma}^{L-Z\gamma \gamma}_{\alpha_1 \alpha_2
\alpha}(k_1,k_2,k)=\sum_{i=1}^3\widetilde{\Gamma}^{Li-Z\gamma
\gamma}_{\alpha_1 \alpha_2 \alpha}(k_1,k_2,k),
\end{equation}
\begin{eqnarray}
\widetilde{\Gamma}^{L1-Z\gamma \gamma}_{\alpha_1 \alpha_2
\alpha}(k_1,k_2,k)&=&k_2\cdot k\epsilon_{\alpha_1 \alpha_2 \alpha
\mu}k^\mu_1-k_1\cdot k\epsilon_{\alpha_1 \alpha_2 \alpha
\mu}k^\mu_2+k_{\alpha_2}\epsilon_{\alpha \alpha_1 \mu
\nu}k^\mu_1k^\nu_2-k_{\alpha_1}\epsilon_{\alpha \alpha_2 \mu
\nu}k^\mu_1k^\nu_2, \\ \widetilde{\Gamma}^{L2-Z\gamma \gamma}_{\alpha_1
\alpha_2 \alpha}(k_1,k_2,k)&=&-k_\alpha \epsilon_{\alpha_1 \alpha_2 \mu
\nu}k^\mu_1k^\nu_2, \\ \widetilde{\Gamma}^{L3-Z\gamma \gamma}_{\alpha_1
\alpha_2 \alpha}(k_1,k_2,k)&=&k_1\cdot k_2\epsilon_{\alpha \alpha_1
\alpha_2 \mu}(k_1-k_2)^\mu+k_{1\alpha_2}\epsilon_{\alpha \alpha_1 \mu
\nu}k^\mu_1k^\nu_2-k_{2\alpha_1}\epsilon_{\alpha \alpha_2 \mu
\nu}k^\mu_1k^\nu_2.
\end{eqnarray}
\end{mathletters}

Note that the vertex functions vanish for three on-shell particles. They are also
symmetric under the interchange of identical particles, in perfect agreement with
Bose symmetry.

\section*{Appendix II: TNGBC vertex functions in the nonlinear
scenario}
\label{a2}

We are using the same conventions used in the linear case. The $CP$-odd
vertex functions can be written in terms of those of the linear scenario
plus some new terms\\

$ZZZ$ vertex:

\begin{mathletters}
\begin{equation}
\Gamma^{NL-ZZZ}_{\alpha_1 \alpha_2
\alpha_3}(k_1,k_2,k_3)=\Gamma^{L-ZZZ}_{\alpha_1 \alpha_2
\alpha_3}(k_1,k_2,k_3)+\sum_{i=3}^5
f^{ZZZ}_{NLi}\Gamma^{NLi-ZZZ}_{\alpha_1 \alpha_2 \alpha_3}(k_1,k_2,k_3),
\end{equation}
\begin{eqnarray}
\Gamma^{NL3-ZZZ}_{\alpha_1 \alpha_2
\alpha_3}(k_1,k_2,k_3)&=&k^2_1(k_{3\alpha_2}g_{\alpha_1
\alpha_3}+k_{2\alpha_3}g_{\alpha_1 \alpha_2})+ \nonumber \\
&&k^2_2(k_{3\alpha_1}g_{\alpha_2 \alpha_3}+k_{1\alpha_3}g_{\alpha_1
\alpha_2})+k^2_3(k_{2\alpha_1}g_{\alpha_2
\alpha_3}+k_{1\alpha_2}g_{\alpha_1 \alpha_3}),\\
\Gamma^{NL4-ZZZ}_{\alpha_1 \alpha_2 \alpha_3}(k_1,k_2,k_3)&=&-k_1\cdot
k_2(k_{3\alpha_3}g_{\alpha_1 \alpha_2}+k_{2\alpha_1}g_{\alpha_2
\alpha_3}+k_{1\alpha_2}g_{\alpha_1 \alpha_3}) \nonumber \\ &&-k_2\cdot
k_3(k_{1\alpha_1}g_{\alpha_2 \alpha_3}+k_{3\alpha_2}g_{\alpha_1
\alpha_3}+k_{2\alpha_3}g_{\alpha_1 \alpha_2}) \nonumber \\ &&-k_1\cdot
k_3(k_{2\alpha_2}g_{\alpha_1 \alpha_3}+k_{3\alpha_1}g_{\alpha_2
\alpha_3}+k_{1\alpha_3}g_{\alpha_1 \alpha_2}), \\
\Gamma^{NL5-ZZZ}_{\alpha_1 \alpha_2
\alpha_3}(k_1,k_2,k_3)&=&k_{3\alpha_2}(k_{3\alpha_1}k_{2\alpha_3}-
k_2\cdot k_3g_{\alpha_1
\alpha_3})+k_{2\alpha_3}(k_{2\alpha_1}k_{3\alpha_2}-k_2\cdot
k_3g_{\alpha_1 \alpha_2})+ \nonumber \\
&&k_{3\alpha_1}(k_{3\alpha_2}k_{2\alpha_3}-k_1\cdot k_3g_{\alpha_2
\alpha_3})+k_{1\alpha_3}(k_{1\alpha_2}k_{3\alpha_1}-k_1\cdot
k_3g_{\alpha_1 \alpha_2})+ \nonumber \\
&&k_{2\alpha_1}(k_{2\alpha_3}k_{1\alpha_2}-k_1\cdot k_2g_{\alpha_2
\alpha_3})+k_{1\alpha_2}(k_{1\alpha_3}k_{2\alpha_1}-k_1\cdot
k_2g_{\alpha_1 \alpha_3}).
\end{eqnarray}
\end{mathletters}

$ZZ\gamma$ vertex:

\begin{mathletters}
\begin{equation}
\Gamma^{NL-ZZ\gamma}_{\alpha_1 \alpha_2
\alpha}(k_1,k_2,k)=\Gamma^{L-ZZ\gamma}_{\alpha_1 \alpha_2
\alpha}(k_1,k_2,k)+\sum_{i=4}^5
f^{ZZ\gamma}_{NLi}\Gamma^{NLi-ZZ\gamma}_{\alpha_1 \alpha_2
\alpha}(k_1,k_2,k),
\end{equation}

\begin{equation}
\Gamma^{NL4-ZZ\gamma}_{\alpha_1 \alpha_2
\alpha}(k_1,k_2,k)=(k_1-k_2)\cdot k (k_{\alpha_2}g_{\alpha_1
\alpha}-k_{\alpha_1}g_{\alpha_2 \alpha}),
\end{equation}
\begin{equation}
\Gamma^{NL5-ZZ\gamma}_{\alpha_1 \alpha_2
\alpha}(k_1,k_2,k)=k_{\alpha_1}(k_{1\alpha}k_{\alpha_2}-k_1\cdot k
g_{\alpha_2 \alpha})+k_{\alpha_2}(k_{2\alpha}k_{\alpha_1}-k_2\cdot k
g_{\alpha_1 \alpha}).
\end{equation}
\end{mathletters}

The respective $CP$-even vertex functions can also been written in terms
of those of the linear case.\\

$ZZZ$ vertex:

\begin{mathletters}
\begin{equation}
\widetilde{\Gamma}^{NL-ZZZ}_{\alpha_1 \alpha_2
\alpha_3}(k_1,k_2,k_3)=\widetilde{\Gamma}^{L-ZZZ}_{\alpha_1 \alpha_2
\alpha_3}(k_1,k_2,k_3)+g^{ZZZ}_{NL3}\widetilde{\Gamma}^{NL3-
ZZZ}_{\alpha_1 \alpha_2 \alpha_3}(k_1,k_2,k_3),
\end{equation}
\begin{eqnarray}
\widetilde{\Gamma}^{NL3-ZZZ}_{\alpha_1 \alpha_2
\alpha_3}(k_1,k_2,k_3)&=&k_1\cdot (k_2-k_3)\epsilon_{\alpha_1 \alpha_2
\alpha_3 \mu}k^\mu_1+\nonumber \\ &&k_2\cdot (k_3-k_1)\epsilon_{\alpha_1
\alpha_2 \alpha_3 \mu}k^\mu_2+k_3\cdot (k_1-k_2)\epsilon_{\alpha_1
\alpha_2 \alpha_3 \mu}k^\mu_3.
\end{eqnarray}
\end{mathletters}

$ZZ\gamma$ vertex:

\begin{mathletters}
\begin{equation}
\widetilde{\Gamma}^{NL-ZZ\gamma}_{\alpha_1 \alpha_2
\alpha}(k_1,k_2,k)=\widetilde{\Gamma}^{L-ZZ\gamma}_{\alpha_1 \alpha_2
\alpha}(k_1,k_2,k)+g^{ZZ\gamma}_{NL6}\widetilde{\Gamma}^{NL6-
ZZZ}_{\alpha_1 \alpha_2 \alpha}(k_1,k_2,k),
\end{equation}
\begin{equation}
\widetilde{\Gamma}^{NL6-ZZ\gamma}_{\alpha_1 \alpha_2
\alpha}(k_1,k_2,k)=k\cdot (k_1-k_2)\epsilon_{\alpha_1 \alpha_2 \alpha
\mu}k^\mu.
\end{equation}
\end{mathletters}

Finally, the $Z\gamma \gamma$ vertex has the same Lorentz structure in both the
linear and the nonlinear scenarios.


\end{document}